\RequirePackage{cmap}
\documentclass[reprint,aps,pra,showemail,showpacs,footinbib]{revtex4-1}
\usepackage[utf8]{inputenc}
\usepackage[english]{babel}
\usepackage{amsthm}
\usepackage{amsmath}
\usepackage{latexsym}
\usepackage{amsfonts}
\usepackage{amssymb}
\usepackage{color}
\usepackage{bbm,dsfont}
\usepackage{graphicx}
\usepackage{enumerate}
\usepackage{hyperref}
\usepackage{subfigure}
\usepackage{tikz}
\usepackage{placeins}
\usepackage[normalem]{ulem}
\newcommand{\stkout}[1]{\ifmmode\text{\sout{\ensuremath{#1}}}\else\sout{#1}\fi}

\newtheorem{proposition?}{Proposition?}

\theoremstyle{definition}

\newcommand{\ket}[1]{|#1\rangle} 
\newcommand{\bra}[1]{\langle#1|} 
\newcommand{\kb}[2]{|#1\rangle\!\langle#2|} 

\newcommand{\fii}{\varphi}

\newcommand{\salg}{\mathcal{F}} 

\newcommand{\diff}{{\rm d}}

\newcommand{\abs}[1]{\vert #1 \vert}

\renewcommand{\phi}{\varphi}
\newcommand{\zpo}{z_{+}^{1}}
\newcommand{\zmo}{z_{-}^{1}}

\newcommand{\azpo}{\abs{z_{+}^{1}}^2}
\newcommand{\azmo}{\abs{z_{-}^{1}}^2}
\newcommand{\azpz}{\abs{z_{+}^{0}}^2}
\newcommand{\azmz}{\abs{z_{-}^{0}}^2}

\begin{document}
\title{Geometric phase gates in dissipative quantum dynamics}
\author{Kai Müller}
\author{Kimmo Luoma}
\email{kimmo.luoma@tu-dresden.de}
\author{Walter T. Strunz}
\affiliation{Institut f{\"u}r Theoretische Physik, Technische Universit{\"a}t Dresden,
D-01062,Dresden, Germany}
\date{\today}
\begin{abstract}
Trapped ions are among the most promising candidates
for performing quantum information processing tasks. Recently,
it was demonstrated how the properties of geometric phases
can be used to implement an entangling
 two qubit phase gate  with significantly reduced operation time
while having a built-in resistance against certain types of errors
{(Palmero et. al., Phys. Rev. A {\bf 95}, 022328 (2017))}.
{In this article, we investigate the influence of both quantum and thermal
  fluctuations on the geometric phase in the Markov regime.}
We show that additional
environmentally induced phases as well as a loss of coherence result from
the non-unitary evolution, {even at zero temperature}.  We
connect these effects to the associated dynamical and geometrical phases. This suggests
a strategy to compensate the detrimental environmental influences and restore
some of the properties of the ideal implementation.
Our main result is a strategy for zero temperature to construct forces for the geometric phase gate
which compensate the
dissipative effects  and leave the produced phase as well as the
final motional state identical to the isolated case. We show that the same strategy
helps also at finite temperatures.
Furthermore, we
examine the effects of dissipation on the fidelity  and the robustness
of a two qubit phase gate against certain error types.
\end{abstract}
\maketitle
\section{Introduction}
    During the last decades there has been an increasing effort to develop
    reliable, large scale quantum information processors. Since such a device
    could utilize quantum properties like superpositions and entanglement, its
    computing power could potentially surpass every conceivable
    classical device for certain problems \cite{Feynman1982,Nielson&Chuang} with potential applications
    in various fields of science and technology.
    At the moment there are several physical realizations
    developed in parallel, each with their own benefits and drawbacks.
    One of the most advanced platforms for quantum information processing is based
    on {trapped ions} \cite{Linke3305}, where many elementary operations have already been
    experimentally demonstrated with high precision \cite{PhysRevLett.113.220501,PhysRevLett.117.060504,PhysRevLett.117.060505}.
    Up to date there are, however, still various
    difficulties to overcome \cite{Wineland1998}. One of the most severe issues
    is dissipation and decoherence  resulting from the interaction of the quantum system
    with the environment leading to detrimental effects on the quantum
    resources and to quantum gate errors.
    Although there
    exist quantum error correction schemes that can compensate small
    errors of the quantum gates these only allow error rates of
    roughly 1\% and come at the cost of a high computational overhead
    \cite{errorrates}. This means that in order to construct  an efficient
    quantum information  processor it is necessary to reduce the error rates of the
    individual quantum gates as much as possible. It is therefore
    important to have a good understanding  of the environmental
    effects and  how to compensate for them.

    In this work we want to
    specifically focus on two-qubit phase gates,
    which perform the following operation
    \begin{align}
    	\label{eq:phaseGateOperation}
    	&\ket{00} \to \ket{00},\hspace{20pt} \ket{11} \to \ket{11},\notag\\
    	&\ket{01} \to e^{i\Phi}\ket{01}, \hspace{6pt}\ket{10} \to e^{i\Phi} \ket{10}.
    \end{align}  
    Two-qubit phase gates are important since
    they can be used to convert the separable state $ 1/2(\ket{11} +
    \ket{10} + \ket{01} + \ket{00} ) $ into a maximally entangled state
    $ 1/2(\ket{11} + i\ket{10} + i\ket{01} + \ket{00} ) $.
     First experimental implementations were
    realized over a decade ago
    \cite{Sackett2000,experimentalDemonstration}, based on theoretical proposals in \cite{PhysRevLett.82.1971,PhysRevA.59.R2539,MSJ}.
    However due to recent efforts in theory \cite{PhysRevA.95.022328,Steane_2014,PhysRevA.71.062309,PhysRevLett.91.157901}
    and experimental techniques \cite{Schafer2018} the operation times and error rates have significantly reduced. These realizations
    leverage the idea of geometric phases first introduced by Berry \cite{Berry45,PhysRevLett.58.1593}
    where the cyclic evolution of a quantum state results in the acquisition of a phase.
    {The aim of this article is to investigate the effects of quantum and thermal
      fluctuations on the geometric phase gate given by Eq.~(\ref{eq:phaseGateOperation}).
      We show how to extend the ideal (fluctuation free) implementation of the gate
    given in~\cite{PhysRevA.95.022328} in order to compensate the
  detrimental environmental effects.}

{The outline of the remainder of the article is the following.}
    In Sec.~\ref{sec:model} we first review the ideal isolated case, introduce
    our notation and then present our open system model in the context
    of a single trapped ion. In Sec.~\ref{sec:geometricPhases} we then show
    how dissipation leads to additional phases
    and in which way they can be connected to the conventional geometrical and
    dynamical phases. Furthermore we show which conditions the
    experimental protocol must satisfy in order to implement a phase gate and how the
    sensitivity of the gate against small experimental errors is
    altered compared to the case where the system is perfectly
    isolated from its environment. In \ref{sec:twoQubit} we then apply our
    results to the two-qubit phase gate protocol proposed in~\cite{PhysRevA.95.022328}
    and examine the impact on the fidelity.
    {In Sec.~\ref{sec:finiteTemperatureEffects} we use the results of the
      previous sections to draw conclusions for the finite temperature case.}
    We close the {main part of the article} with a summary and an outlook.
    {Lastly, some of the analytic computations are presented in the Appendix.}

\section{Model}
    \label{sec:model}
    In this section we will first introduce a model {for}  an isolated
    phase gate. Then we
    expand the model to account for
    {detrimental environmental effects.}

\subsection{Isolated system}
We consider the ion trap as a quantum harmonic oscillator with mass $m$
and frequency $\omega$  which is driven by an
external force leading to the Hamiltonian
    \begin{equation}
    	\label{eq:modelIsolated}
    	H_{{\mathrm{isol}}}(t) = \hbar\omega a^{\dagger}a + V(t).
    \end{equation}
    Here $a$ $(a^\dagger)$ is the annihilation (creation) operator for the vibrational
    mode satisfying bosonic commutation relations $[a,a^\dagger]=1$. The potential $ V(t) $ arises
    from the externally applied force $ F(t) $. Since we want to implement the operation
    described in Eq.~\eqref{eq:phaseGateOperation} we need to introduce
    state-dependent forces $ F_{1} $ and $ F_{0} $ that depend on an
    internal (e.g. spin-) state of the ion in order to distinguish these states. An ion in the internal state
    $ \ket{1} $ will only experience $ F_{1} $ and vice versa. In
    the following we will use the notation $ F_{j} $ with $j = 0,1$ labeling
    the internal state of the ion. Furthermore, the external forces $ F_{j}(t) $ are assumed to be
    homogeneous over the extent of the motional state. This can be assumed, for example,
    for forces realized by lasers if
    the wavelength of the laser is much greater than the amplitude of
    oscillation. Under
    these circumstances the Hamiltonian can be written in the following
    form
    \begin{align}
        \label{eq:Hschroedinger}
        H(t) &= \hbar\omega(a^{\dagger}a) + \kb{0}{0}\otimes V_{0}(t) + \kb{1}{1}\otimes V_{1}(t), \\
        V_{j}(t) &= F_{j}(t) x = f_j(t)(a+a^{\dagger}),
    \end{align}
    with $ f_j(t) = \dfrac{\hbar}{2m\omega}F_j(t) $.

    Before we determine the evolution of a quantum state in this model we will
    simplify the equations more by switching to an interaction picture with
    respect to $\hbar\omega a^\dagger a$ leading to a simpler Hamiltonian
     \begin{align}
    \label{eq:coupledPotential}
    \widetilde{H}(t) &= \kb{0}{0}\otimes \widetilde{V}_{0}(t) + \kb{1}{1}\otimes \widetilde{V}_{1}(t),\\
    \widetilde{V}_{j}(t) &= \widetilde{f}_j^{*}(t)a+\widetilde{f}_j(t)a^{\dagger}, \notag
    \end{align}
    where $\widetilde{f}_{j} = e^{i\omega t} f_{j}$.
    The equation of motion in the interaction picture
    for a quantum state represented by the density operator $ \rho $
    is the von-Neumann equation
    \begin{equation}
    \label{eq:VNisolated}
    \dot{\rho} = -i\frac{1}{\hbar} [\widetilde{H}(t),\rho].
    \end{equation}
    This equation can be solved by inserting
    {an ansatz $\rho=\kb{\Psi_t}{\Psi_t}$}, where
    \begin{align}
    \label{eq:ansatzisol}
      {\ket{\Psi_t}} =& {\sum_{j=0}^1a_je^{i\fii_j(t)}\ket{j,z_j(t)}},
    \end{align}
    where $ j $ represents the internal state and
    {$\ket{z_j(t)}=e^{-\abs{z_j(t)}^2/2}\sum_{n=0}^\infty ((z_j(t))^n/\sqrt{n!})\ket{n}$ is a coherent state
      for the motional degree of freedom of the ion when the internal state is  $\ket{j}$, ($j=0,1$)
    and $a_j$ is a constant determined from the initial conditions.}
    Inserting this ansatz into Eq.~(\ref{eq:VNisolated}) leads to the following equation
    for the coherent state label  $z_j(t)$
    \begin{equation}
    \label{eq:z(t)isol}
    \dot{z}_{j} = \dfrac{1}{i\hbar}\widetilde{f}_{j}(t).
    \end{equation}

    {In the {context of implementing} a phase gate{,} we want $ f_j(t) $ to be part of some protocol
    which is switched on at a certain time and is completed some time $ T $ later.
    Therefore $ f_j $ shall only be non-zero in the interval $ \left[0,T\right] $ and it shall be such
    that the motional state undergoes a cyclic evolution $ z_{j}(0) = z_{j}(T) $
    whereas the internal degrees of freedom acquire a phase according to Eqs.~(\ref{eq:phaseGateOperation}).}
    It is known that such a cyclic quantum evolution
    leads to the acquisition of a phase $\phi_j = \phi_{\mathrm{g},j}+\phi_{\mathrm{d},j}$, where the dynamical and
    geometrical phases satisfy $\dot\phi_{\mathrm{d},j}=-(1/\hbar)\bra{j,z_j(t)}H(t)\ket{j,z_j(t)}$
    and $\dot\phi_{\mathrm{g},j}=i\bra{j,z_j(t)}\partial_t\ket{j,z_j(t)}$, respectively.
    The total phase acquired  is equal to twice the area enclosed by the trajectory $ z_j(t) $
    in the interaction picture (see Eq.~(\ref{eq:phiA}))~\cite{experimentalDemonstration,PhysRevA.95.022328}.
    In the following we will expand this model to include
    {quantum and thermal fluctuations}.

\subsection{Open system}
\label{ssec:dissipativeCase}
Effects of the ion coupling
  to some external environment
are modeled phenomenologically by a
Gorini-Kossakowski-Sudarshan-Lindblad (GKSL)
{master} equation \cite{Lindblad1976,doi:10.1063/1.522979} of the following
form
{\begin{align}
    \label{eq:model}
    \dot{\rho} &= -i\frac{1}{\hbar} [\widetilde{H}(t),\rho] + \gamma(\bar{n} +1) (2a\rho a^{\dagger} - a^{\dagger}a\rho - \rho a^{\dagger}a)\notag\\
               & + \gamma\bar{n} (2a^{\dagger}\rho a - aa^{\dagger}\rho - \rho aa^{\dagger}),
  \end{align}}{in the interaction picture}.
   {The model (\ref{eq:model}) describes dissipation
      and thermal excitation of the motional state of the ion with rates
      $\gamma(\bar{n} +1)$ and $\gamma\bar{n}$, respectively.  $\bar{n}$ models the
      average occupation number of the bosonic heat bath modes at a relevant system
      frequency at finite temperature. At zero temperature, $\bar{n}=0$ and only quantum fluctuations
      and damping with rate $\gamma$ are present.
      }

      {{Motional} coherence times $(\gamma\bar{n})^{-1}$ for the trapped ion systems are
      of the order of {1 - 100 milliseconds~\cite{Wineland1998, lucas2007longlived}}. Typical frequency
      for the harmonic motion of the ion around the minimum of the trap is in the
      MHz range, whereas operation times for the two qubit phase gate investigated
      later in the article is in the $\mu$s range~\cite{PhysRevA.95.022328}.}

      {
        As shown in \cite{PhysRevA.74.022102,Strunz2005},
        finite temperature effects can be incorporated in the zero temperature
        model by adding a fluctuating force $\sqrt{2\gamma\bar{n}}\hbar\chi(t)$
        to the Hamiltonian.
        Here $\chi(t)$ is a Gaussian white noise process  with
        $\langle\chi(t)\rangle = \langle \chi(t)\chi(t')\rangle = 0 $ and
        $\langle\chi(t)\chi^*(t')\rangle = \delta(t-t')  $.
        Thus the Hamiltonian reads
    \begin{align}
   \label{eq:H_xi}
   \widetilde{H_{\chi}}(t) &= \kb{0}{0}\otimes \widetilde{V}_{0,\chi}(t) + \kb{1}{1}\otimes \widetilde{V}_{1,\chi}(t),\\
   \widetilde{V}_{j}(t) &= (\widetilde{f}_j^{*}(t) + \sqrt{2\gamma\bar{n}}\hbar\chi^*(t))a\\
   &+(\widetilde{f}_j(t)+ \sqrt{2\gamma\bar{n}}\hbar\chi(t))a^{\dagger}. \notag
   \end{align}
   Note that the noise does not depend on the internal state $j$.
   The finite temperature model~(\ref{eq:model}) is thus equivalent to
   an ensemble of zero temperature models  with stochastic Hamiltonian
   \begin{align}
   \label{eq:rho_xi}
   \dot{\rho_{\chi}} = -i&\frac{1}{\hbar} [\widetilde{H_{\chi}}(t),\rho_{\chi}] +
                           \gamma (2a\rho_{\chi} a^{\dagger} - a^{\dagger}a\rho_{\chi} - \rho_{\chi} a^{\dagger}a).
   \end{align}
   The evolution of the density operator can be recovered by taking an average over
   $\chi(t)$
   \begin{equation}
   \label{eq:rhoAverage}
   \rho(t) = \langle\rho_{\chi}(t)\rangle,
   \end{equation}
   and the average state $\rho(t)$ satisfies Eq.~(\ref{eq:model}).}

 {Remarkably, Eq.~(\ref{eq:rho_xi}) can still be solved
   by a coherent state ansatz, such as Eq.~(\ref{eq:ansatzisol}),
   although it contains the effects of thermal and quantum fluctuations.
   Inserting the ansatz from Eq.~\eqref{eq:ansatzisol}
   for a particle in the internal state $\ket{j}$ 
   into Eq.~(\ref{eq:model}) leads to the following equation
   for the coherent state label  $z_j(t)$
   \begin{equation}
     \label{eq:z(t)}
     \dot{z}_{j} + \gamma z_{j} = \dfrac{1}{i\hbar}\left(\widetilde{f}_{j}(t) + \sqrt{2\gamma\bar{n}}\hbar\chi(t)\right).
   \end{equation}}
   {For current ion-traps the motional coherence time $1/(\gamma\bar{n})$ is much longer
   than the operation time~\cite{Wineland1998, lucas2007longlived}} so that the motion
     of the ion is dominated by the deterministic force.\\
     In the following sections we will first investigate the zero temperature  case
     for an arbitrary force. In section \ref{sec:finiteTemperatureEffects} we can apply
     these results to a noisy force and average over the noise.
\section{Consequences of quantum fluctuations}
    \label{sec:geometricPhases}
In this section we investigate the consequences of coupling the
trapped ion to a zero temperature bath. The effect of thermal
fluctuations will be considered in Sec.~\ref{sec:finiteTemperatureEffects}.
\subsection{Consequences for the phase}
    \label{ssec:consequncesPhase}
    In the following we want to investigate how the Lindblad terms in the
    time evolution equation affect the phase. We therefore consider a
    model which is in principle identical to (\ref{eq:model})
    {at zero temperature} but slightly
    more general. The result can then be applied to the phase gate model.

\begin{figure}
    	\includegraphics[width= 0.7\linewidth]{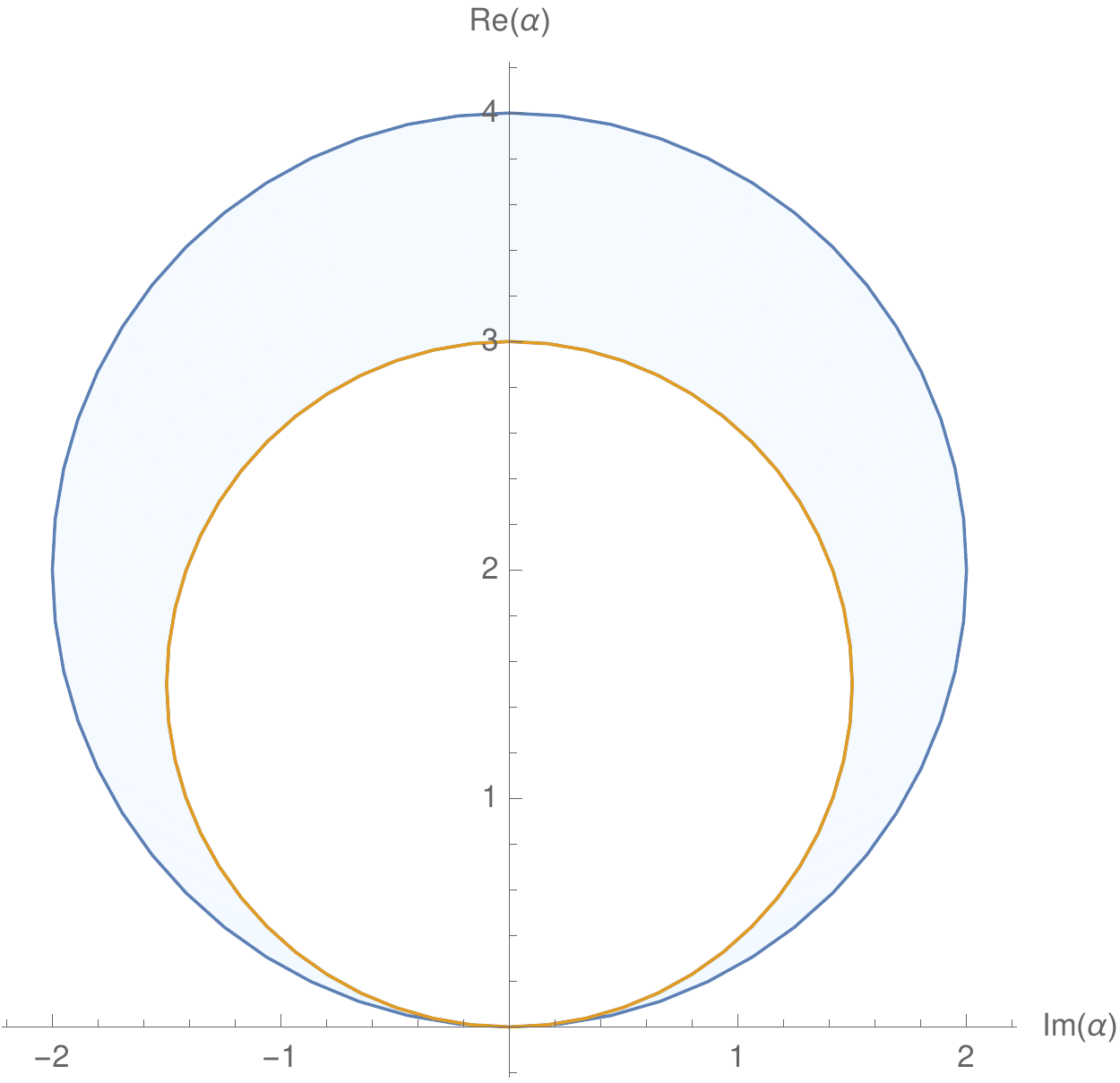}
    	\caption{(color online) In the isolated case the relative phase between two cyclic
          evolutions is proportional to the difference of the swept phase space
          area in the interaction picture.}
    	\label{fig:phasespacearea}
    \end{figure}

    The Hamiltonian shall be of the form
    \begin{equation*}
    	H(t) = \kb{0}{0}\otimes H_0(t) + \kb{1}{1}\otimes H_1(t),
    \end{equation*}
    where $ \ket{0} $, $ \ket{1} $ represent the internal states of
    the ion and $ H_0(t) $ and $ H_1(t) $ act on the motional degree of freedom.
    We assume that the dissipation and decoherence is well described by a
    general GKSL master equation and thus arrive at the following model
    \begin{align}
    	\label{eq:time evolution}
    	\dot{\rho} &= \frac{-i}{\hbar}[H(t),\rho] + \mathcal{L}\left[\rho\right],\\
    	\mathcal{L}[\rho] &= \sum_{l=1}^{N} L_l\rho L_l^{\dagger} - \dfrac{1}{2}\left(L_l^{\dagger}L_l\rho + \rho L_l^{\dagger}L_l\right).\notag
    \end{align}
    Furthermore we assume that this model has a pure state solution
   {$\kb{\Phi_t}{\Phi_t}$,
      where $\ket{\Phi_t}=\sum_j\ket{j,\Psi_j(t)}$ and the}
     internal state $ j $ remains unchanged during the evolution. Although these are very limiting
    assumptions we will see that the results can nevertheless be applied to
    the phase gate scenario mentioned before.\\
    Under these assumptions, we can repeat the argument proposed in~\cite{PhysRevLett.58.1593}
    for a cyclic quantum evolution governed by a
    time-dependent Schrödinger equation to determine the relative phase that arises if the initial state is in a superposition of internal states. In our case, however, the cyclic quantum
    evolution of a pure state is modified by a damping term. The details
    of the computation can be found in the appendix~\ref{sec:Anhangphidissipative}.
    We then get a new complex valued term $\xi$ in addition to the dynamical and geometrical phase
    \begin{flalign}
    \label{eq:phi}
    	\phi &= \phi_{\mathrm{g}} + \phi_{\mathrm{d}} + \xi,\\
    	\xi &= \phi_{\mathrm{L}} + i\eta,
    \end{flalign}
    where the individual terms are defined as follows:
    \begin{align}
    \dot{\phi}_{\mathrm{g}} &= i\left(\bra{\Psi_0}\partial_t\ket{\Psi_0} - \bra{\Psi_1}\partial_t\ket{\Psi_1}\right), \notag\\
    	\dot{\phi}_{\mathrm{d}} &= -\frac{1}{\hbar}(\langle H_0(t)\rangle - \langle H_1(t)\rangle),\\
    	\dot{\xi} &= -i \sum_{l=1}^{N}\bra{\Psi_0}L_l\ket{\Psi_0}\bra{\Psi_1}L_l^{\dagger}\ket{\Psi_1} \notag\\
    	  &- \dfrac{1}{2}\left(\bra{\Psi_0}L_l^{\dagger}L_l\ket{\Psi_0} + \bra{\Psi_1}L_l^{\dagger}L_l\ket{\Psi_1}\right). \notag
    \end{align}
    The first two terms are identical to the unitary case mentioned before and
    also found in~\cite{PhysRevLett.58.1593}.
    Therefore, they correspond to dynamical and
    geometrical phases which arise during a cyclic evolution of a quantum
    system. Since we have constructed relative phases they are expressed
    as the difference between the dynamical/geometrical phases of
    particles in the internal states $0$ and $1$.
    The last sum cannot be expressed in such a way and
    contains dissipative effects.  In general it leads to real terms in the
    exponent which result in a loss of coherence. We can apply this equation to the
    damped harmonic oscillator if we set $ L = \sqrt{\gamma}a $ and $
    \ket{\Psi_j} = \ket{j,z_j} $.  Furthermore we can identify $H_j(t)$, which
    corresponds to the Hamiltonian seen by a particle in the internal state
    $\ket{j}$ as $\hbar\omega a^{\dagger}a + V_j(t)$ (see
    Eq.~(\ref{eq:Hschroedinger})). This means we can calculate the dynamical
    phase for a particle in the internal state $j$ with $\ket{j,z_j}$ in
    the interaction picture by using Eqs.~(\ref{eq:z(t)}) and~\eqref{eq:coupledPotential} as
    \begin{align}
      \phi_{\mathrm{d},j} &= -\frac{1}{\hbar}\langle H_j(t)\rangle \notag\\
        &= \frac{-1}{\hbar}\int\limits_{0}^{T}\bra{j,z_j(t)} e^{-i\omega t a^{\dagger}a} H(t) e^{i\omega t a^{\dagger}a}\ket{j,z_j(t)} \diff t\notag\\
    		&= \frac{-1}{\hbar}\int\limits_{0}^{T}\bra{j,z_j(t)}\hbar\omega a^{\dagger}a + \widetilde{V}_j(t)\ket{j,z_j(t)} \diff t\notag\\
    		&= \int\limits_{0}^{T}2\operatorname{Im}\left(\dot{z}_j(t) z_j^{*}(t)\right) - \omega\abs{z_j(t)}^2 \diff{t}.
    \end{align}
    For the geometric phase we arrive at
    \begin{flalign}
      \phi_{\mathrm{g},j} &= i\left(\bra{z_j(t)}e^{-i\omega t a^{\dagger}a}\partial_te^{i\omega t a^{\dagger}a}\ket{z_j(t)}\right) \notag\\
        &= \int\limits_{0}^{T}-\operatorname{Im}(\dot{z}_j(t)z_j^{*}(t)) + \omega\abs{z_j(t)}^2 \diff{t},\\
    \end{flalign}
    again with $\ket{j, z_j}$ in the interaction picture. As we can see
    the dynamical and geometrical phase are remarkably similar for the
    harmonic oscillator. Furthermore we can combine these two phases for
    the total phase in the isolated ($\gamma = 0$) case $\phi_{{\mathrm{isol}}}$:
    \begin{flalign}
      \phi_{{\mathrm{isol}}} &= (\phi_{d,0} - \phi_{d,1}) + (\phi_{g,0} -\phi_{g,1}) \notag\\
      \label{eq:phaseAsIm}
        &= \int\limits_{0}^{T}\operatorname{Im}\left((\dot{z}_0(t)z_0^{*}(t)\right) - \dot{z}_1(t)z_1^{*}(t)) \diff{t}.
    \end{flalign}
    For a cyclic evolution this reduces to the known result
    \cite{PhysRevA.95.022328,experimentalDemonstration}
    \begin{equation}
    \label{eq:phiA}
     \phi_{{\mathrm{isol}}} = 2 (A_0 - A_1),
    \end{equation}
    where $A_j$ is the area enclosed by the cyclic evolution of
    $z_j$. This is shown in Fig.~\ref{fig:phasespacearea}. From now on,
    we do not always write the time dependence of the coherent state labels
    explicitly in order to shorten the notation.

    The influence of the dissipation  is {contained}
    in the term $ \xi $ with
    {
    \begin{align*}
    \label{eq:phiL}
    	i\dot{\xi} &= \gamma\left(2z_0z_1^{*} - \left(\abs{z_0}^2 + \abs{z_1}^2\right)\right)\\
    		&= -\gamma \abs{z_{1}-z_{0}}^2  + i \gamma\abs{z_{1}}\abs{z_{0}}\sin(\theta_{1} - \theta_{0}),
    \end{align*}
    }where the phases $\theta_j$ are defined by $z_j=\abs{z_j}e^{i\theta_j}$.
    {Note that $ \xi $ consists of a real as well as of an imaginary part.\\
    In summary, an initial state which is in a superposition of spin states
    {\begin{equation*}
    	\rho(0) =
    	\begin{pmatrix}
    	\abs{a}^{2} && ab^{*} \\
    	a^{*}b && \abs{b}^{2}
    	\end{pmatrix} \otimes \kb{z(0)}{z(0)},
    \end{equation*}}
    will be transformed into the following state after the cyclic evolution:
    {\begin{equation*}
    	\rho(T) =
    	\begin{pmatrix}
    	\abs{a}^{2} && ab^{*} e^{i\phi_{{\mathrm{isol}}} + i\xi} \\
    	a^{*}b e^{-i\phi_{{\mathrm{isol}}} - i\xi^{*}} && \abs{b}^{2}
    	\end{pmatrix} \otimes \kb{z(0)}{z(0)}.
    \end{equation*}}
    This shows that for $ \gamma \neq 0 $ the damping results in an
    additional real term $ \eta = \gamma\int_{0}^{T} \abs{z_{1}-z_{0}}^2 \diff t $ in the exponent which does only depend on the
    damping strength $ \gamma $ and the difference of the amplitudes of the paths.  This real
    term leads to a dephasing of the spin state by diminishing the off-diagonal elements of the density matrix.
    We will therefore refer to it as dephasing term from now on. }

    We can also see a new phase term
    $\phi_{\mathrm{L}}= \gamma\int\limits_{0}^{T}\abs{z_{1}}\abs{z_{0}}\sin(\theta_{1} - \theta_{0}) \diff t$
    which depends on the absolute values of  $ z_0 $
    and $ z_1 $.  The integral over this term can vanish for sufficiently
    symmetrical $ z_j(t) $ (e.g. if $ z_j(t) = z_j(T-t) $) with $
    j\in\{0,1\} $ or if $ z_{1} $ or $ z_{0} $ is in the ground state
    during the entire operation.  In Sec.~\ref{sec:twoQubit}
    we will see that the 2-qubit phase gates proposed in \cite{PhysRevA.95.022328,PhysRevA.71.062309}
    {and realized in \cite{experimentalDemonstration}} do indeed
    have the latter property which means that even with damping the
    phases produced by those phase gates are still only determined by the
    respective areas. The dephasing term can, however, only vanish if $
    f_{1} = f_{0} $ which implies that there is no relative phase as
    well. We can also conclude that the dephasing is stronger for higher
    energies of the particle which means it is especially relevant for
    short operation times as we will see in Sec.~\ref{ssec:fidelity}.

\subsection{Consequences for the path}
    \label{ssec:force}
    We have seen in the previous section how the damping results in
    additional phase terms. However, from Eq.~(\ref{eq:z(t)}) it is clear that
     the damping alters the path as well.  Therefore, the paths which are
    closed in the isolated case are no longer closed in the damped case.
    It is a natural question to ask which forces $ \widetilde{f}_j(t) =
    f_j(t)e^{i\omega t} $ can be used to achieve the cyclical
    evolution ($z_j(0) = z_j(T) $) in the damped case and whether some of
    those forces should be used preferably because they minimize the
    dephasing term. First we note that it is not possible to completely
    compensate the effects of the damping by applying some sophisticated
    force $f_{\mathrm{c}}$. This can be seen from Eq.~(\ref{eq:z(t)}), since the isolated
    dynamics of a coherent state is described by $\dot{z}_j = 0$, such a
    force would need to satisfy $\widetilde{f} = e^{i\omega t}f_{\mathrm{c}}(t) = \mathrm{const} $,
    which is impossible for real $f_{\mathrm{c}}(t)$.

    To determine
    the effects of a force $ f_j(t) $ on the path $ z_j(t) $ we have to
    solve Eq.~(\ref{eq:z(t)}).
    {This leads to the solution}
    \begin{flalign}
    	z_j(t) &= z_{j,\mathrm{hom}} + z_{j,\mathrm{inhom}} \notag\\
    	\label{eq:alpha(t)}
    		&= z_j(0) e^{-\gamma t} + \int_{0}^{t} \dfrac{-i}{\hbar} \widetilde{f}_j e^{-\gamma (t-\tau)} \diff{\tau}.
    \end{flalign}
    We can therefore conclude that in order to achieve the cyclic dynamics
    $ z_j(0) = z_j(T) $ we need the forces to satisfy
    \begin{equation}
    	\label{eq:conditionf}
    	z_j(0) \left(e^{\gamma T} - 1\right) = \int_{0}^{T}f_j(\tau)e^{i\omega\tau}e^{\gamma\tau} \diff{\tau}.
    \end{equation}
    The equation shows explicitly that for $ \gamma \neq 0 $ the condition
    depends on the initial state $ z_j(0) $. This means that in contrast to
    the undamped case where $ f_j $ would always lead to closed trajectories it now
    only works for a specific initial condition $z_j(0)$.
    The fault tolerance of a quantum phase gate
    towards the initial motional state is therefore lost in the damped
    case. \\
    An interesting observation at this point is that if we
    consider $z_j(0) = 0$, we can derive forces $ f_{\mathrm{d}} $ which
    return $ z_j $ to the ground state after time $ T $ in the damped
    case  from the forces $ f_{\mathrm{nd}} $ which accomplish this in
    the undamped case  by using the formula
    \begin{equation}
    \label{eq:relationFDamping}
    	f_{\mathrm{d}} = f_{\mathrm{nd}}\cdot e^{-\gamma t}.
    \end{equation}
    We will use this link between the damped and the undamped scenarios in the
    next section to generalize an already existing protocol for 2-qubit
    phase gate to account for dissipative effects.

     For an experimental realization
    it is desirable to minimize the dephasing term for a given relative
    phase. To examine how this can be done we want to consider the case
    $f_{0}(t) = 0 $ and $ \ket{z_j(0)} = \ket{0} $ for the sake of
    simplicity. This means that $ \ket{z_0} $ is the ground
    state at all times and we only need to discuss the dynamics
    of $ \ket{z_1} $.  These simplifications are well justified because in
    an experimental setup the ground state can be prepared initially and
    an additional force $ f_{0} $ does not bring any benefits but just
    makes the computations more complex.  We can then derive simple
    expressions for the phase and dephasing terms after time $ T $
    from Eq. \eqref{eq:phiA} if we write $ z_1 = r(t) e^{i\theta(t)} $
    {(in the following equations we have omitted the time dependence of
    $ r $ and $ \theta $)}
    \begin{flalign*}
    	\int_{0}^{T} z_1^{*}\dot{z_1} \diff{t} &= \int_{0}^{T} re^{-i\theta}\left(\dot{r} e^{i\theta} + ir\dot{\theta} e^{i\theta}\right)\diff t \\
    		&= \int_{0}^{T} r\dot{r} \diff{t} + i\int_{0}^{T} \dot{\theta}r^2 \diff{t}.
    \end{flalign*}
    Since $ z_1(0) = z_1(T) $ we can see that the first integral vanishes
    by integrating by parts and we are left with the following expression
    for the phase:
    \begin{equation}
    \label{eq:comparisonPhase}
    	\phi_{\mathrm{isol}}= \int_{0}^{T} \dot{\theta}r^2 \diff{t},\quad \phi_{\mathrm{L}}=0.
    \end{equation}
    In this case, $\phi_{\mathrm{L}}=0$ because $z_0=0$ at all times.
    For the dephasing term we find
    \begin{equation}
    \label{eq:comparisonDephasing}
    	\eta=\gamma \int_{0}^{T} r^2 \diff{t}.
    \end{equation}
    We can see that the easiest way to minimize
    the dephasing for a given relative phase is to make $ \dot{\theta} $
    as large as possible. This result also makes intuitive sense since if
    the path goes around the origin multiple times (large $ \dot{\theta} $)
    it needs a smaller amplitude (which results in less dephasing) in
    order to sweep over the same area. Since $\theta$ corresponds to the
    interaction picture it is affected by the frequency $\omega$ of
    the harmonic oscillator.

\begin{figure}
      \centering
    		\includegraphics[width = \linewidth]{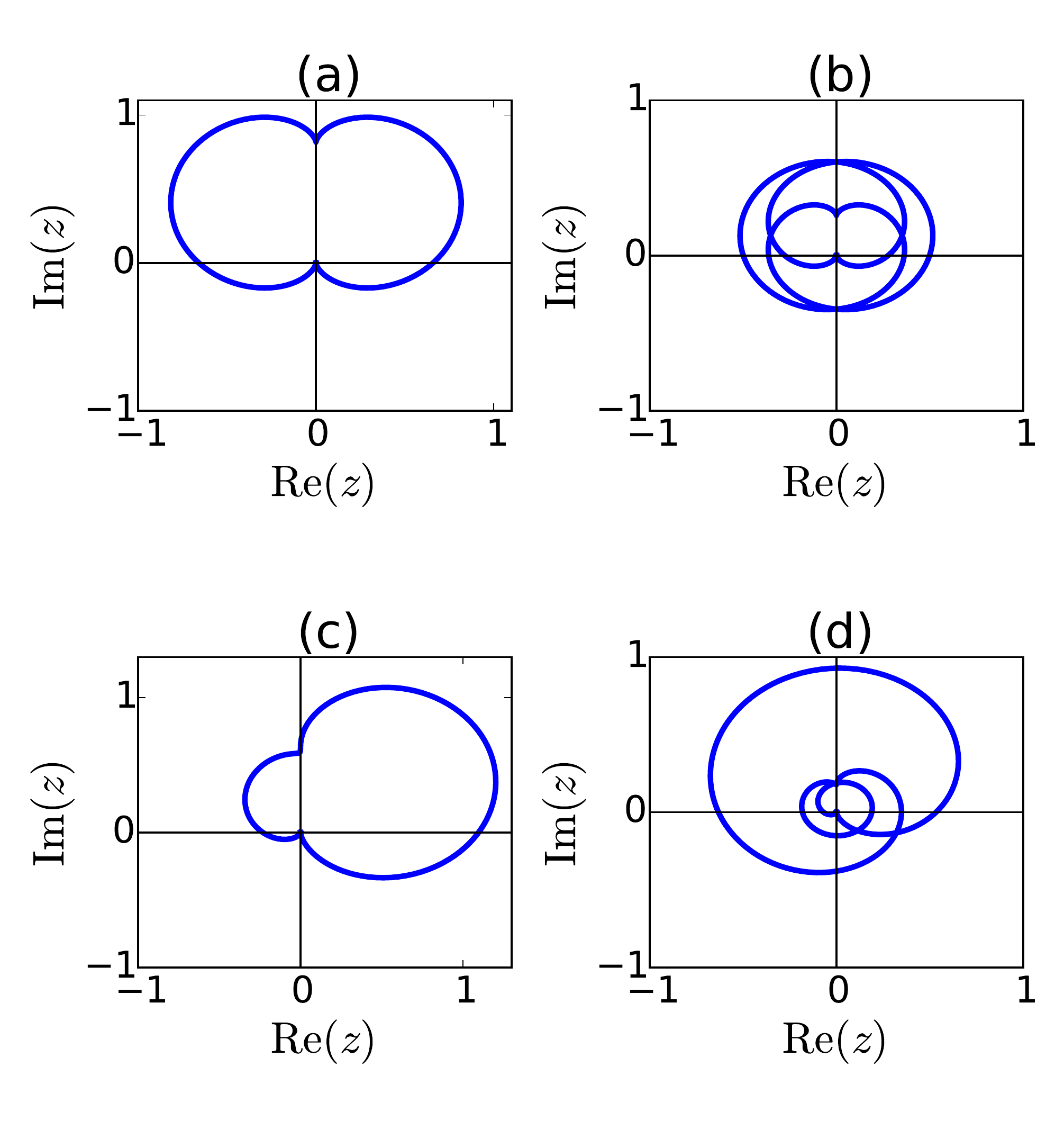}
    	\caption{(color online){Path of $ z_1 $ in the complex plane
            {when using the force $ f(t) = e^{-\gamma t} \sin(\Omega t) $}.
            All paths were normalized to produce a phase of $ \pi $. The
            paths in the upper row (a),b)) were simulated without damping,
            whereas damping was included for the paths in the lower row (c),d))
            without changing the frequency of the trap or the driving force.
            The paths in the right column are simulated in a trap with a
            higher frequency and result in less dephasing although they
            produce the same phase compared to the left column.}}
    		\label{fig:paths1}
    \end{figure}

    Fig.~\ref{fig:paths1} displays the paths
    $z_1(t)$ which are generated by forces of the form $f_1 =
    e^{-\gamma t} \sin(\Omega t)$ and $f_{0} =0$
    (see Eq.~(\ref{eq:z(t)})).
    The paths (a) and (b) correspond to the isolated case with $\gamma =
    0$ whereas $\gamma > 0$ for the paths (c) and (d).
    {The exact parameters used in this numerical simulation were
      $ \Omega = 2\,\mathrm{MHz}$ for the frequency  of the driving force in all trajectories,
      $ \omega = 2\Omega $ for the frequency of the harmonic trap in
      trajectories (a) and (c) and $ \omega = 4\Omega $ for (b) and (d).
      The damping constant for paths (c) and (d) was set to $\gamma/\Omega = 0.2$.}
    We can see how
    the upper two paths are symmetrical with respect to the imaginary
    axis. As shown in~\cite{PhysRevA.95.022328} this implies that the
    phase does not change (in first order) if the force is subjected to a homogeneous,
    small constant offset $ f \mapsto f + \delta f $ in the $ \gamma = 0 $
    case. In contrast, the paths for $ \gamma \neq 0 $ are no longer
    symmetrical.
    However whether the path is symmetric or not depends on the
    force. In the next section we therefore want to investigate how
    the robustness can be maintained in the damped case by
    constructing forces differently compared to Eq.~\eqref{eq:relationFDamping}.
\subsection{Consequences for the robustness}
    \label{sec:consequenceRobustness}
    At first we want to show how the condition for the robustness against
    small constant offsets of the force $ f \mapsto f+\delta f $ reads in the
    dissipative case studied here. According to Eq.~(\ref{eq:phaseAsIm}) with $
    z_0 = 0 $,
    \begin{flalign*}
    	\phi_{\mathrm{isol}} &= \int_{0}^{T}\operatorname{Im}(\dot{z}_1 z_1^{*}) \diff t \notag\\
    		&= \dfrac{-1}{\hbar}\int_{0}^{T}\operatorname{Re}(z_{1}^{S} f_1) \diff t,
    \end{flalign*}
    where we used Eq.~\eqref{eq:z(t)} and $ z_{1}^{S} = e^{-i\omega t} z_1 $ is
    the path in the Schrödinger picture. Therefore the offset to the phase
    in first order becomes
    \begin{equation*}
    	\delta \phi_{\mathrm{isol}} = \dfrac{-\delta f}{\hbar} \int_{0}^{T}\operatorname{Re}(z_{1}^{S}) \diff t \overset{!}{=} 0.
    \end{equation*}
    This result is identical to the one in the isolated case found in
    \cite{PhysRevA.95.022328}. By inserting Eq.~\eqref{eq:alpha(t)}, assuming $
    z_{1}(0) = 0 $ and integrating by parts we can express this as a
    condition for the force
    \begin{equation}
    	\label{eq:forceResisanceCondition}
    	0 = \int_{0}^{T}f(t)\diff t.
    \end{equation}
    Together with the condition for a cyclic evolution, Eq.~\eqref{eq:conditionf},
    we therefore have a set of conditions that for $ z_j(0) = 0 $ may be
    seen as orthogonality conditions for $ f(t) $
    \begin{equation}
    	\label{eq:orthogonal}
    	f(t) \perp \{e^{\gamma t} \sin(\omega t), e^{\gamma t} \cos(\omega t), 1\} =: \mathcal{C}.
    \end{equation}
    This means that we can construct forces to suit our needs by a
    Gram-Schmidt procedure. It is useful to orthogonalize the set $
    \mathcal{C}$ and then do one more orthogonalizing step for the
    arbitrary function $g$ which will then become orthogonal
    to the set $ \mathcal{C} $. This method makes it possible to construct
    a plethora of forces which will leave the phase unchanged under a
    small constant offset $ \delta f $ and produce a cyclic evolution. By
    superposing many of such forces one can then ensure to meet
    further demands like e.g. $ f(0) = f(T) = 0 $.

    We can conclude that
    it is possible to maintain the robustness of the phase gate against
    small constant offsets of the force $ f \mapsto f +\delta f $ in the
    damped case. However as we have seen in the previous section (Eq.~\eqref{eq:conditionf}) the gate
    loses its resistance against fluctuations in the initial motional
    state.
\section{Application to 2-qubit phase gates}
    \label{sec:twoQubit}
    In this section we want to show how the relations we found in the
    previous sections apply to two-qubit phase gates which have been realized in
    \cite{experimentalDemonstration,Schafer2018}.
    These two-qubit gates consist of two
    ions in a harmonic trap potential which experience a force that
    depends on the internal state ($ \ket{\uparrow} $ or $
    \ket{\downarrow} $) of the ion. As shown in \cite{PhysRevA.95.022328},
    the Hamiltonian of such a system can be written as
    \begin{align}
    	H_{\mathrm{tot}} &= H_{+} + H_{-},\notag\\
    	\label{eq:H2Qubit}
    	H_{\pm} &= \dfrac{p_{\pm}^2}{2} + \dfrac{1}{2}\Omega_{\pm}^2 x_{\pm}^2 + f_{\pm} x_{\pm}.
    \end{align}
    Here, $ H_{+} $ describes an oscillation of a stretch mode where the
    displacement from the equilibrium position of the two ions are equal but
    in opposite directions and $ H_{-} $ describes an oscillation of the
    center-of-mass mode where the displacement of the ions is
    identical. {$ x_{\pm} $ are mass weighted normal mode
      coordinates which for equal mass ions take the form
    \begin{equation*}
    		x_{\pm} = \sqrt{m}((x_{1}-x_{1}^{(0)}) \mp (x_{2}-x_{2}^{(0)})).
    \end{equation*}
    Here, $ x_{i} $ is the position operator for  ion $ i $, $ x_{i}^{(0)} $ is the equilibrium
    position of ion $i$ and the canonically conjugate momentum operators
    are $p_{\pm}=-i\hbar\partial/\partial x_{\pm} $.}

    Note that
    here we ignored a term in $ H_{\mathrm{tot}} $ which is proportional to the
    difference of the forces experienced by the two ions. This (purely time dependent)
    term will therefore lead to additional phases for certain configurations. Similar to Eq.~\eqref{eq:orthogonal} we
    will however later (see Eq.(\ref{eq:orthogonality2qubit})) present a way to
    construct forces which satisfy $ \int_{0}^{T} f \diff t = 0 $ so that this phase will vanish.
    A more detailed derivation of the Hamiltonian and discussion of the purely time dependent
    term can be found in \cite{PhysRevA.95.022328}.

    {If the forces on the two ions take the form $ F_j = F(t)
    \sigma_{j}^{z} $ one can derive the following values for the force in the interaction picture $ \widetilde{f}_{\pm} $
    \begin{flalign}
    	\widetilde{f}_{+}(P) &= \widetilde{f}_{-}(A) = 0, \notag\\
    	\widetilde{f}_{-}(\uparrow\uparrow) &= -2F/\sqrt{2m}e^{i\Omega_- t},\notag\\
    	\widetilde{f}_{+}(\uparrow\downarrow) &= -2F/\sqrt{2m}e^{i\Omega_+ t},\notag\\
    	\widetilde{f}_{-}(\downarrow\downarrow) &= 2F/\sqrt{2m} e^{i\Omega_- t}, \notag\\
    	\label{eq:2Qubitf}
    	\widetilde{f}_{+}(\downarrow\uparrow) &= 2F/\sqrt{2m} e^{i\Omega_+ t},
    \end{flalign}
    where $ P \in \{\uparrow\uparrow, \downarrow\downarrow\} $ denotes
    parallel and $ A\in \{\uparrow\downarrow, \downarrow\uparrow\} $
    anti-parallel spin combinations.  This type of force can be realized
    by off resonant lasers in the Lamb-Dicke regime~\cite{schleich2011quantum}.
    For the optimization of the functional form of the
    force see \cite{PhysRevA.95.022328}. Frequencies $\Omega_\pm$ are
    defined as $\Omega_+ = \sqrt{3}\omega$ and $\Omega_-=\omega$~\cite{PhysRevA.95.022328}.} We can bring the Hamiltonian in the
    same form as in the previous section by introducing
    creation and annihilation operators for the stretch and center-of-mass
    mode $ a_{\pm}, a_{\pm}^{\dagger} $ and switching to an interaction
    picture
    \begin{flalign}
    	\ket{\Psi_{I}} &= e^{-iH_{0}t/\hbar}\ket{\Psi}, \notag\\
    	\label{eq:2QubitInteractionPicture}
    	H_{0} &= \hbar\Omega_{+}(a_{+}^{\dagger}a_{+} + \dfrac{1}{2}) + \hbar\Omega_{-}(a_{-}^{\dagger}a_{-} + \dfrac{1}{2}).
    \end{flalign}
    The Hamiltonian then reduces to
    \begin{flalign}
    	\label{eq:2QubitInteractionHamiltonian}
    	\widetilde{V} &= \widetilde{f}_{+}^{*}a_{+} + \widetilde{f}_{+}a_{+}^{\dagger} + \widetilde{f}_{-}^{*}a_{-} + \widetilde{f}_{-}a_{-}^{\dagger} \notag\\
    		&= \widetilde{V}_{+} + \widetilde{V}_{-}.
    \end{flalign}
    To model the damping we can introduce two Lindblad terms similar to
    Eq. \eqref{eq:model}.
    We assume identical damping rates $\gamma$ since all degrees of freedom couple
    to the same bath which we assume to have a flat spectral density on the
    relevant frequency scale. We then find in the interaction picture
    \begin{align}
    	\label{eq:2QubitTimeEvolution}
    	\dot{\rho} &= \mathcal{L}_{+}[\rho] + \mathcal{L}_{-}[\rho] ,\\
    	\mathcal{L}_{\pm} &:= \frac{-i}{\hbar}[\widetilde{V}_{\pm},{\rho}]
                            + \gamma\left(2a_{\pm}\rho a_{\pm}^{\dagger} - a_{\pm}^{\dagger}a_{\pm}\rho - \rho a_{\pm}^{\dagger}a_{\pm}\right). \notag
    \end{align}
    We can see that the $ \mathcal{L}_{\pm} $ only act on one of the two
    modes and are of the same form  as the right hand side of Eq.~(\ref{eq:model}).
    Therefore the two modes $+,-$ are not coupled by Eq.~\eqref{eq:2QubitTimeEvolution} and we can reuse the results from the previous Sec.~\ref{sec:geometricPhases} to determine the evolution of an initial state that is in a superposition of spin states and in a coherent motional state
    \begin{equation*}
    	\rho(0) =
    	\begin{pmatrix}
    	\abs{a}^{2} && ab^{*} \\
    	a^{*}b && \abs{b}^{2}
    	\end{pmatrix} \otimes \kb{z_{+}(0)}{z_+(0)}\otimes\kb{z_{-}(0)}{z_{-}(0)}.
    \end{equation*}
    The coherent motional state evolves according to
    \begin{equation}
    	\label{eq:2Qubitalphat}
    	\dot{z}_{\pm}^{j}+\gamma z_{\pm}^{j} = \dfrac{1}{i\hbar}\widetilde{f}_{\pm}(j),
    \end{equation}
    where the internal degrees of freedom are given by $j$ and oscillation
    in the stretch and center-of-mass mode is described by
    $z_{+}^{j}$ and $z_{-}^{j}$ respectively. The forces $\widetilde{f}_{\pm}(j)$ are given in Eq.~\eqref{eq:2Qubitf}.
    Furthermore, we can observe that Eq.~\eqref{eq:2QubitTimeEvolution} is of the
    form of Eq.~\eqref{eq:time evolution} which means that we can apply the
    results from Sec.~\ref{ssec:consequncesPhase} to calculate the
    phase and dephasing which arise during a cyclic evolution:
    \begin{equation}
    \label{eq:2QubitA}
    	\phi(T) = \phi_{{\mathrm{isol}}} + \xi,
    \end{equation}
    where
    \begin{align}
    	\phi_{{\mathrm{isol}}} &= 2(A_{+}^{1} + A_{-}^{1} - A_{+}^{0} - A_{-}^{0}),\notag\\
    	\xi &= \gamma \int_{0}^{T} d_{+}(\tau) + d_{-}(\tau) \diff{\tau}, \notag\\
    	d_{\pm} &=  i\abs{z_{\pm}^{1}-z_{\pm}^{0}}^2 + \abs{z_{\pm}^0}\abs{z_{\pm}^1}\sin(\theta_{\pm}^1 - \theta_{\pm}^0).\notag
    \end{align}
    We see that $\phi_{{\mathrm{isol}}} = \phi_{\mathrm{g}} + \phi_{\mathrm{d}}$ is proportional to the areas
    swept in the stretch and center-of-mass modes of oscillation $A_{+}$ and
    $A_{-}$, respectively,  and it is identical to the phase of the isolated evolution
    ($\gamma =0$). Because of the symmetries of the forces described in
    Eq.~\eqref{eq:2Qubitf}, $\phi_{{\mathrm{isol}}}$ is only nonzero if $ 0 $ is a parallel and $
    1 $ an anti-parallel spin combination or vice versa. The $\xi$ term
    originates from the Lindblad operators and since $\operatorname{Im}(\xi) \neq 0$
    it will result in dephasing equivalently to the one ion case discussed
    in Sec.~\ref{ssec:consequncesPhase}. However, if the evolution starts in the
    ground state either $z_{\pm}^{0}$ or $z_{\pm}^{1}$ will
    remain in the ground state for the entire operation because of
    Eq.~\eqref{eq:2Qubitf}. Therefore $\operatorname{Re}(\xi)$ will always be zero which
    means that the phase depends only on the difference of the swept phase
    space areas like in the undamped case studied in
    \cite{PhysRevA.95.022328}.

    Since the equations \eqref{eq:2Qubitalphat} for the evolution of $ z $
    are identical to the one ion case,
    Eq. \eqref{eq:relationFDamping} still holds true and we can easily
    generalize the force $ F_{\mathrm{nd}}$ constructed in
    \cite{PhysRevA.95.022328} to the damped oscillator:
    \begin{align}
    \label{eq:f_damped}
    	F(t) = &\kappa e^{-\gamma t}\cdot F_{\mathrm{nd}}. 
    \end{align}
    According to Eq.~(\ref{eq:alpha(t)}) this means for the damped path
    \begin{equation*}
        {z_{\mathrm{d}}=\kappa e^{-\gamma t}z_{\mathrm{nd}}.}
    \end{equation*}
    Here we introduced two correction factors $ \kappa $ and
   ${e^{-\gamma t}}$ to the original force for the undamped case. $
    \kappa $ is a constant to compensate for the smaller area due to the
    damping and it therefore ensures that the phase (which corresponds to
    the area) stays the same. The exponential factor ensures that $ z
    $ returns to the ground state after time $ T $.

\begin{figure}
    		\includegraphics[width=.6\linewidth]{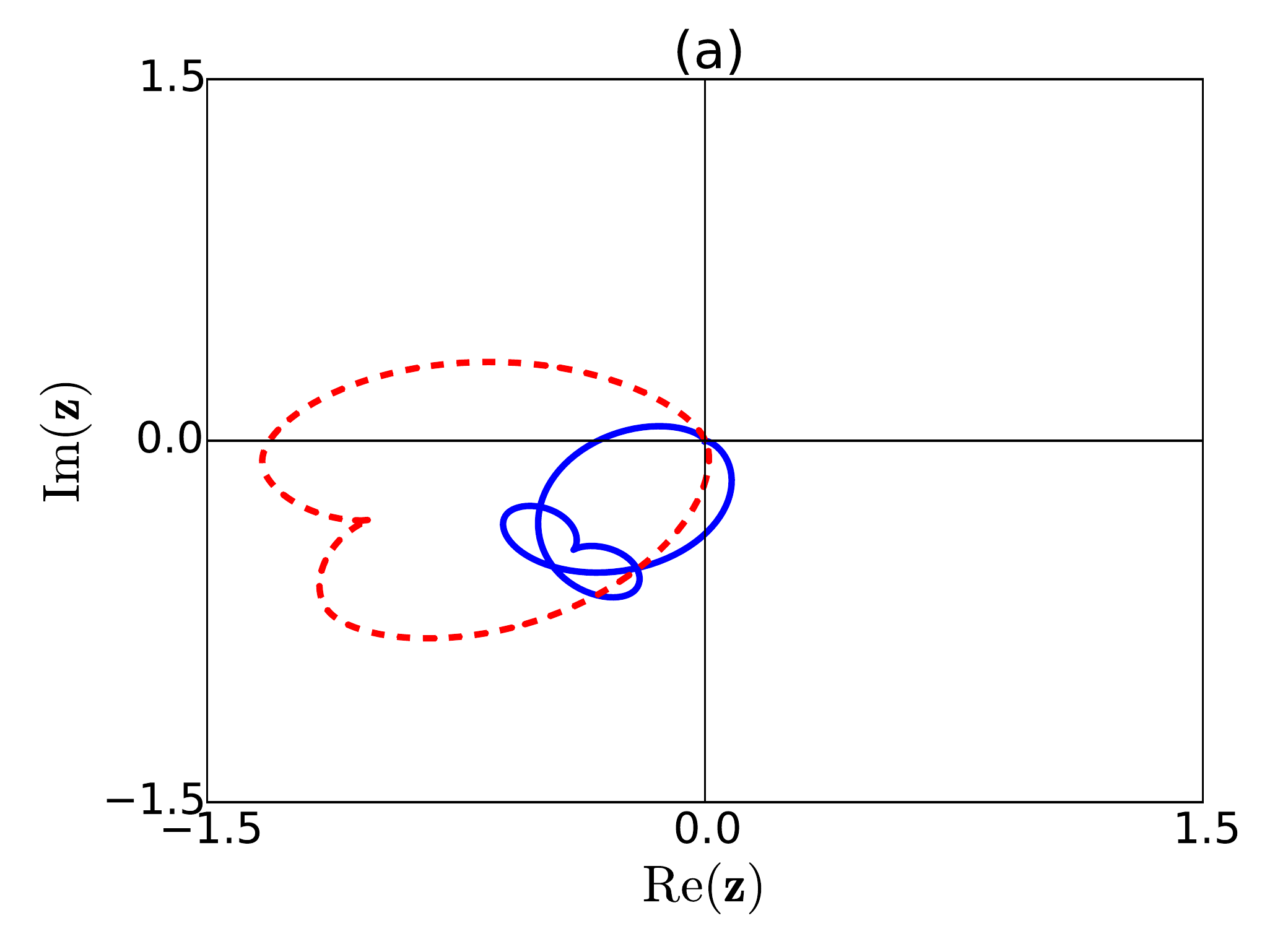}
    		\includegraphics[width=.6\linewidth]{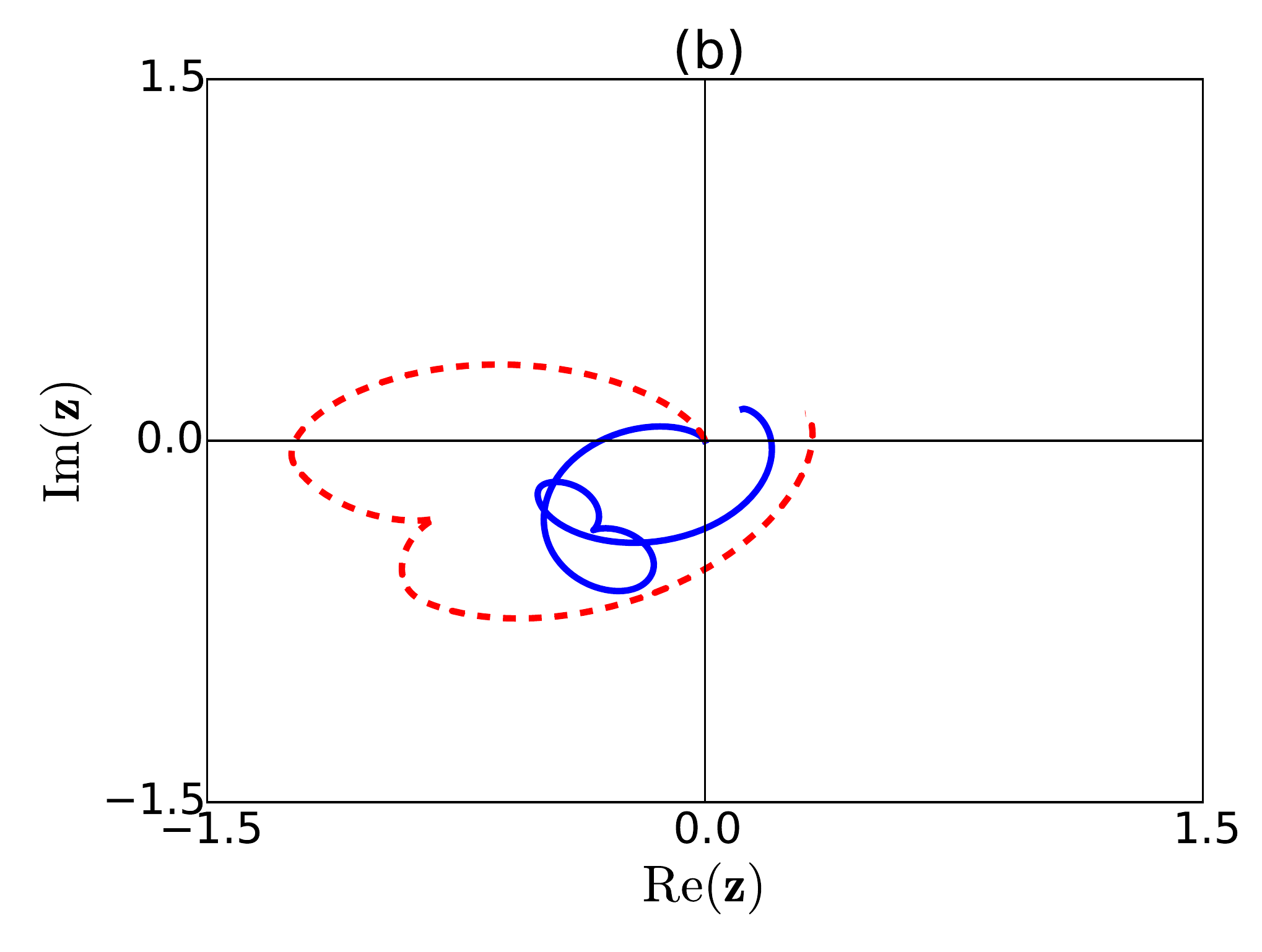}
    		\includegraphics[width=.6\linewidth]{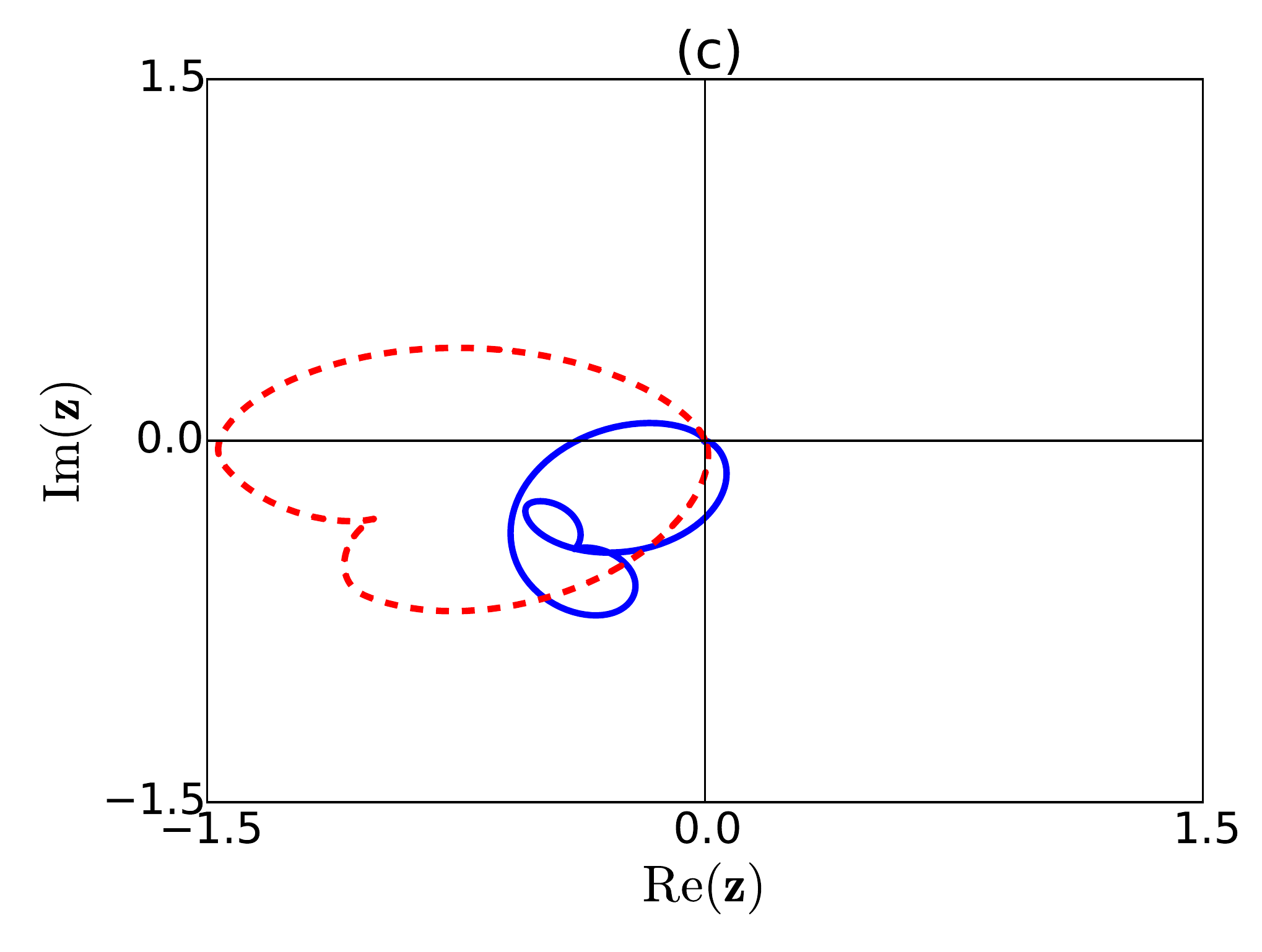}
    		\includegraphics[width=.6\linewidth]{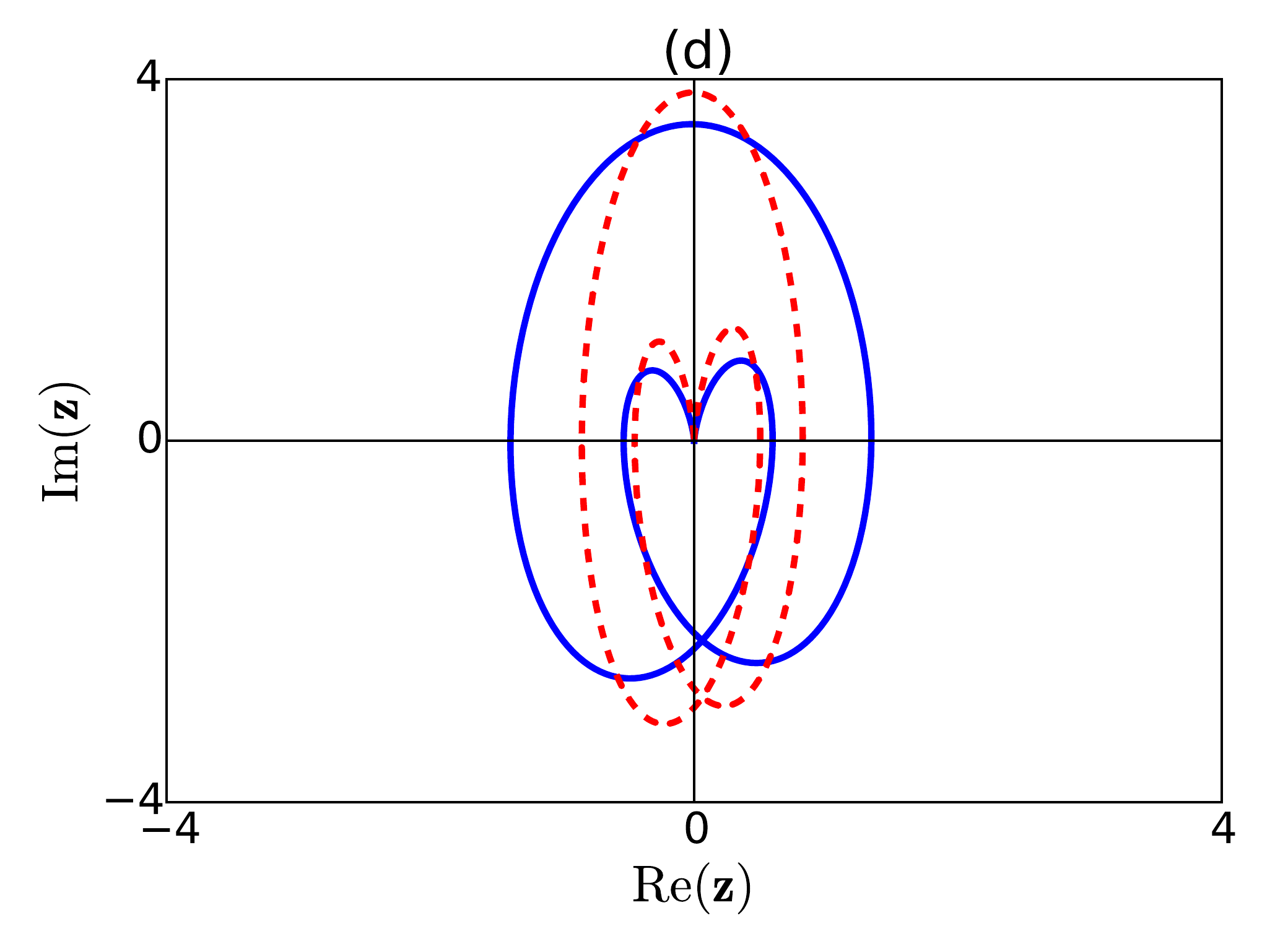}
    	\caption{(color online) Figure (a), (b) and (c) show paths of $
    z_{+}(\uparrow\downarrow) $ (solid blue line) and $ z_{-}(\uparrow\uparrow)
    $ (dashed red line) in the interaction picture. Figure (d) shows
    paths in the Schrödinger picture. Plot (a) corresponds to one of the
    cases studied by \cite{PhysRevA.95.022328}. The damping was set to
    zero and the other parameters were chosen as $ T =
    {0.8}{\mu}$s, $ \omega/(2\pi) = {2}$ MHz.
    Plot (b) shows a path generated by the same force but in the
    presence of fairly strong damping $ \gamma = 0.1\omega $. We can
    clearly see how the paths are no longer closed which would lead to a
    loss of fidelity. This can be fixed by applying the force \eqref{eq:f_damped}
    which accounts for the damping as shown in (c).
    Plot (d) shows the trajectory for a shorter operation time
     $ T ={0.3}{\mu}$s in the Schrödinger picture. Note that this trajectory has a much
    greater amplitude which leads to more dephasing. Because of that, we can conclude that shorter
    operation times come with the trade-off of more dephasing (see
    Fig.~\ref{fig:Infidelity}).}
    	\label{fig:2QubitAlphaPath}
    \end{figure}

    It would also be
    possible to construct forces via the Gram-Schmidt process described in
    Sec.~\ref{sec:consequenceRobustness} to maintain the resistance against small
    constant offsets of the force. Since these forces now have to produce
    closed paths in both modes there are more orthogonality conditions
    \begin{align}
    	\label{eq:orthogonality2qubit}
    	f \perp& \{e^{\gamma t} \sin(\Omega_+ t), e^{\gamma t} \cos(\Omega_+ t), \notag\\
        &\quad e^{\gamma t} \sin(\Omega_{-} t), e^{\gamma t} \cos(\Omega_{-} t), 1\},
    \end{align}
    but the method of constructing $ f $ stays the same. For the next
    section we will nevertheless stick to Eq.~(\ref{eq:f_damped}) since the
    resulting forces are less complex and sufficient for discussing the
    impact on the fidelity. \\
    The paths of the resulting $ z_{\pm} $ with
    and without damping and under the influence of different forces are
    shown in Fig.~\ref{fig:2QubitAlphaPath}. The plots (a), (b) and (c) are in
    the interaction picture whereas the plot  (d) is in the Schrödinger
    picture. The parameters for trajectory (a) were chosen identically to
    \cite{PhysRevA.95.022328}: $ T = {0.8}$ $\mu$s,
    $\omega/2\pi = {2}$ MHz and $ \gamma/\omega = 0 $. The
    trajectory (b) corresponds to the same force and parameters as
    trajectory (a) but now with a damping $ \gamma/\omega =
    0.1 $. We can see how the path is no longer closed and that the area
    decreased as well. In the plot (c) we used the same damping and
    parameters as in (b) but with the adjusted force
    \eqref{eq:f_damped}. Now the paths are closed again and the area
    difference is identical to (a). The plot  (d) shows the trajectory for
    a shorter operation time $ T
    = {0.3}{\mu}$s in the Schrödinger picture. It is
    important to note that the ticks are different for this plot because
    the trajectory has a much greater amplitude. The intuitive reason for
    this is that the particle needs a large momentum to complete the loop
    in a shorter time, the trajectory is therefore stretched out in $ p
    \propto \operatorname{Im}(z^{S}) $ direction. This illustrates that
    shorter operation times come with the trade-off of more dephasing (see
    Sec.~\ref{ssec:fidelity} and Fig.~\ref{fig:Infidelity}).
\subsection{Influence of damping on the fidelity}
    \label{ssec:fidelity}
\begin{figure}
    	\includegraphics[width= \linewidth]{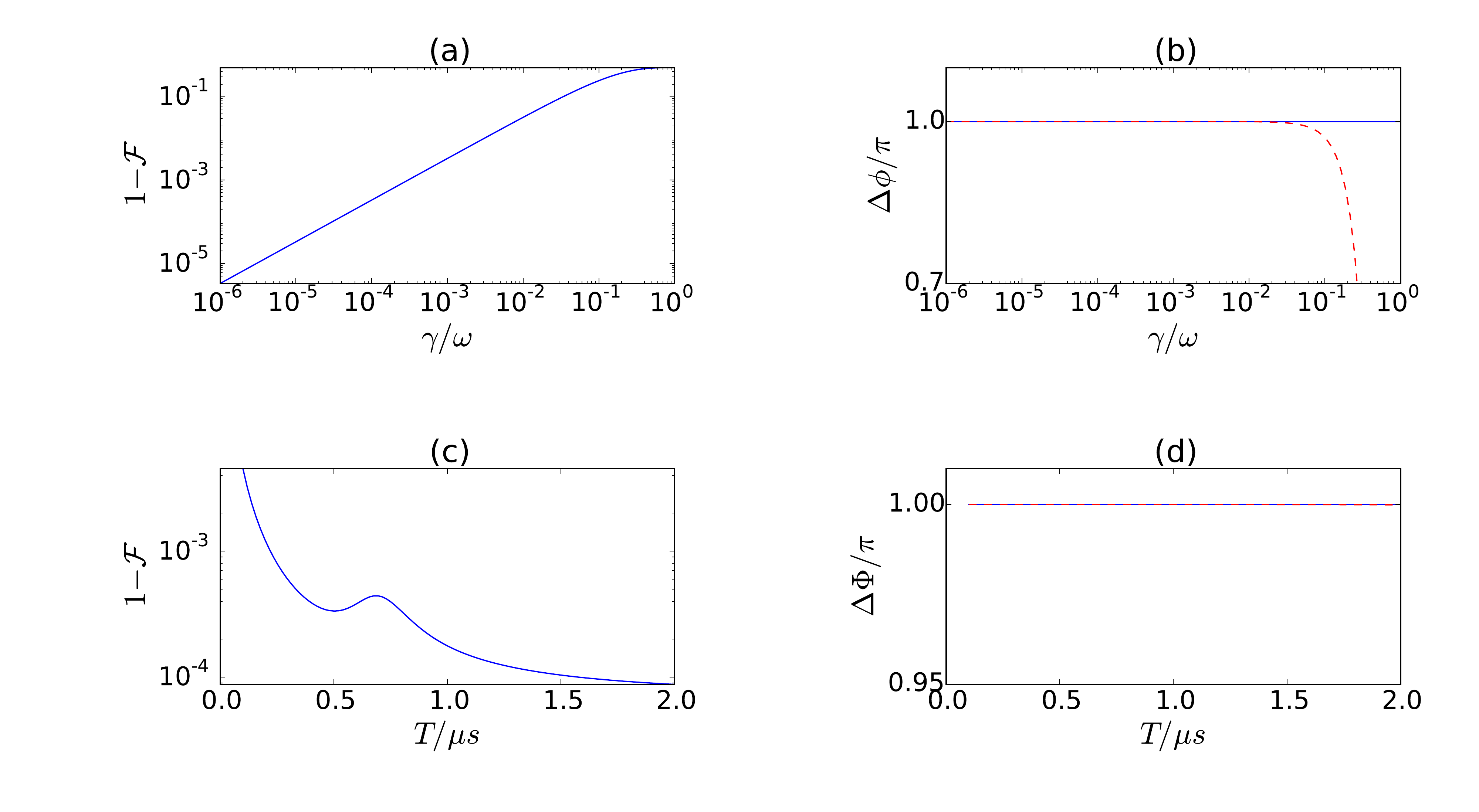}
    	\caption{(color online) The upper two plots show the maximum infidelity (a) and
    phase (b) $ \Delta\phi = 2(\phi(A) - \phi(P)) $ for different values of $
    \gamma $ with constant $ T={0.8}{\mu}$s and the plots (c)
    and (d) show these values for different operation times and constant $
    \gamma/\omega = 10^{-4} $. The solid blue line in (b) and (d)
    represent the phase difference of trajectories which result from the
    force \eqref{eq:f_damped} which accounts for the damping and the
    dashed red lines represent the phase difference of trajectories which
    result from the original force.}
    	\label{fig:Infidelity}
    \end{figure}

    The fidelity measures the overlap of the final state $ \rho_{\mathrm{f}} $ with the desired state $ \ket{\Psi_{\mathrm{d}}} $
    \begin{equation*}
    	\mathcal{F} = \bra{\Psi_{\mathrm{d}}}\rho_{\mathrm{f}}\ket{\Psi_{\mathrm{d}}}.
    \end{equation*}
    {Since we want to implement a} two qubit phase gate our desired state $ \ket{\Psi_{\mathrm{d}}} $ is
    \begin{equation*}
    	\ket{\Psi_{\mathrm{d}}} = ae^{i\phi_{{\mathrm{isol}}}}\ket{P} + b\ket{A},
    \end{equation*}
    with $ \abs{a}^2 + \abs{b}^2 = 1 $. $P$ and $A$  denote an
    arbitrary parallel and anti-parallel spin
    combination(e.g. $ P =\uparrow\uparrow $ and $ A=\uparrow\downarrow $). The final (spin-) state (in the basis $ \{ \ket{P}, \ket{A}\} $)
    after the cyclic evolution is
    \begin{equation*}
    	\rho_{\mathrm{f}} = \begin{pmatrix}
    	\abs{a}^2 & ab^{*} e^{i\phi_{{\mathrm{isol}}} - \Gamma} \\
    	a^{*}b e^{-i\phi_{{\mathrm{isol}}} -\Gamma} & \abs{b}^2
    	\end{pmatrix},
    \end{equation*}
    where
    \begin{equation}
    \label{eq:Gamma}
    	\Gamma = \gamma\int_{0}^{T} \abs{z_{+}(A)}^2 + \abs{z_{-}(P)}^2 \diff{t}.
    \end{equation}
    We can calculate the fidelity as
    \begin{equation}
    	\label{eq:fidelity}
    	\mathcal{F} \geq \dfrac{1 + e^{-\Gamma}}{2}.
    \end{equation}
    The lower bound of the inequality above is reached if the prefactors satisfy $a=b=1/\sqrt{2}$.\\
    Figure \ref{fig:Infidelity} shows the maximal infidelity $ 1 -
    \mathcal{F} $ and phase difference $ \Delta\phi = 2(\phi(A) - \phi(P))
    $ for different $ \gamma $ and $ T $. In (a) and (b) the operation
    time was chosen as constant $ T={0.8}{\mu}$s whereas $
    \gamma/\omega $ varied and in (c) and (d) we chose constant $ \gamma/\omega =
    10^{-4} $ for varying $ T $. The plots (b) and (d) show
    the phase difference for the force given in Eq.~\eqref{eq:f_damped} which accounts
    for the damping (solid blue line) and for the original force (dashed
    red line). We can see that the force we constructed (blue line)
    correctly compensates for the phase whereas lack of  compensation would lead
    to a drastic phase deviation for larger damping strengths. Different
    operation times do not affect the phase significantly for both forces
    because for $ \gamma/\omega = 10^{-4} $ the lost area is still
    marginal. For large $ \gamma $ the curve in (a) goes towards $ 1/2 $
    apart from that the infidelity is roughly in the same order of
    magnitude as $ \gamma/\omega $. In the plot (c) it is demonstrated
    that short operation times $ T $ lead to a higher infidelity, because
    the forces needed to achieve the desired phase result in a higher
    amplitude $ \abs{z} $ and therefore in a larger $ \Gamma $ (see
    Fig.~\ref{fig:2QubitAlphaPath} (d)). The bump in figure (c) comes from our
    choice of the path and has no other physical meaning. We can compare these
    infidelities which are roughly of the order of $ \gamma/\omega $ to
    infidelities from different sources studied by
    \cite{PhysRevA.95.022328}. The infidelity caused by the anharmonic
    Coulomb repulsion is below $ 10^{-4} $ whereas the infidelity caused
    by considering the correct sinusoidal form of the force $ F(x,t) =
    F(t)\cdot\sin(kx) $ is in between $ 10^{-5} $ and $ 0.1 $ depending on
    the operation time.
\section{Finite temperature effects}
    \label{sec:finiteTemperatureEffects}
\begin{figure}
      \includegraphics[width=0.5\textwidth]{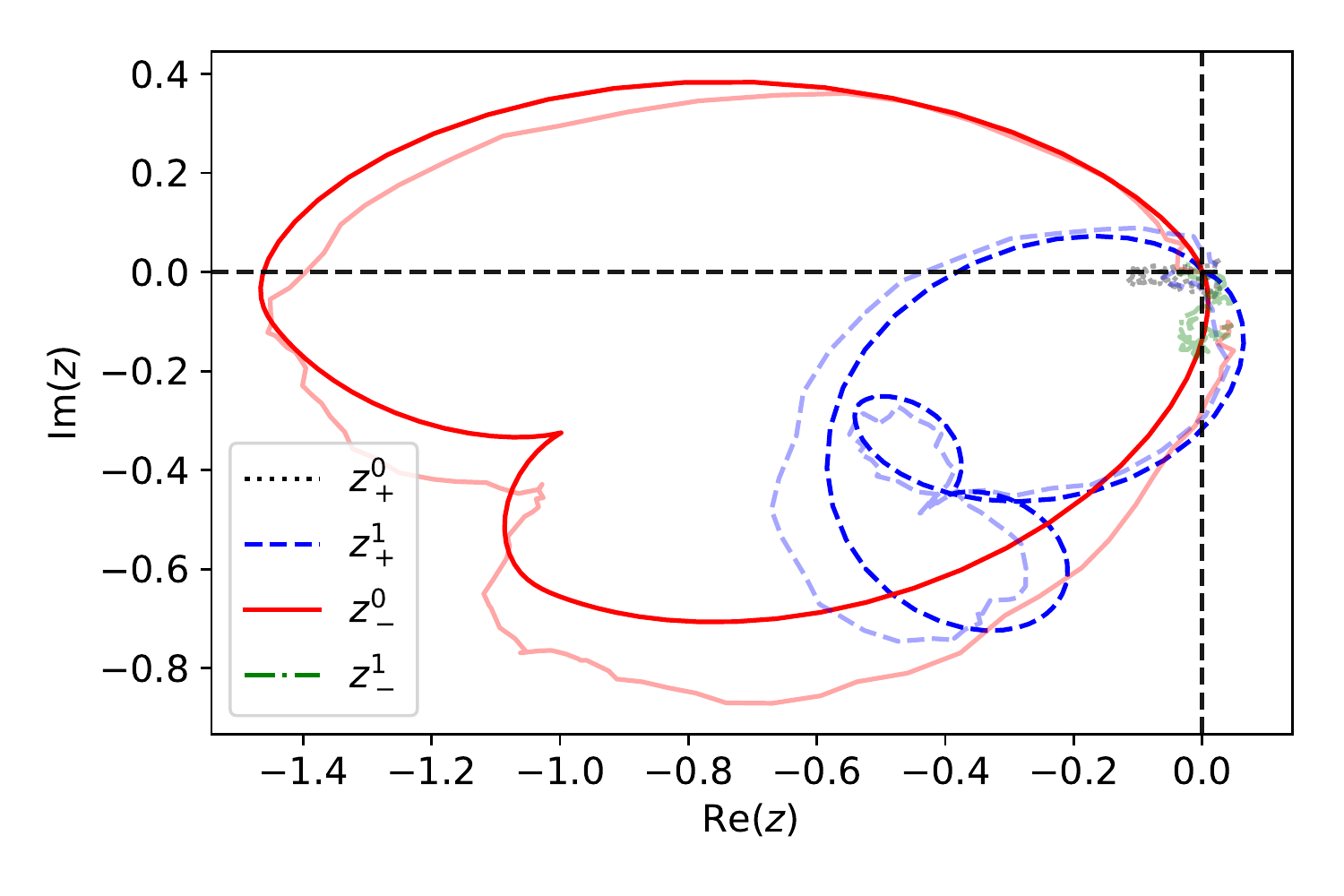}
      \caption{\label{fig:thermal_traj}{(color online) Comparison of
      trajectories between finite temperature ($\gamma\bar{n}T\approx 0.03$) (light colors)
    and zero temperature (dark colors). Thermal fluctuations are clearly visible
    around the zero temperature paths $z_+^1,\,z_-^0$.  Moreover  $z_+^0, z_-^1$ fluctuate around
    the ground state $(0,0)$ when temperature is non-zero, leading to imperfections
    in the phase gate implementation. Other parameters are as in Fig.~\ref{fig:2QubitAlphaPath} c).}}
    \end{figure}

    In section \ref{ssec:dissipativeCase} we outlined how the effects of
      finite temperature can be taken into account by using a noisy force equal to
      $ \widetilde{f}_{j}(t) + \sqrt{2\gamma\bar{n}}\hbar \chi(t) $, where $ \chi(t) $
      is a complex valued Gaussian white noise process.
      In this section we show how the previous results can be extended to include
      finite temperature effects and what impact these effects have on the fidelity
      of a two-qubit phase gate.\\
      We begin with a stochastic GKSL-type master equation that describes the influence of a
      finite temperature heat bath coupled to the trapped ions implementing
      the two qubit phase gate. We model the effect of the heat bath by equal strength but independent
    damping of both vibration modes ($a_\pm$) and by independent thermal
    noise processes affecting both modes ($\chi_\pm(t)$)
    \begin{align}
        \label{eq:2QubitTimeEvolutionFiniteT}
        \dot{\rho_{\chi}} =& \mathcal{L}_{\chi,+}[\rho_\chi] + \mathcal{L}_{\chi,-}[\rho_\chi] ,\\
        \mathcal{L}_{\chi,\pm}[\rho] =& \frac{-i}{\hbar}[\widetilde{V}_{\chi,\pm},{\rho}]
                            + \gamma\left(2a_{\pm}\rho a_{\pm}^{\dagger} - a_{\pm}^{\dagger}a_{\pm}\rho - \rho a_{\pm}^{\dagger}a_{\pm}\right). \notag\\
        \widetilde{V}_{\chi,\pm} =& (\widetilde{f}_{\pm}^{*}+ \sqrt{2\gamma\bar{n}}\hbar \chi_\pm^{*}(t))a_{\pm}
                                    + (\widetilde{f}_{+}+ \sqrt{2\gamma\bar{n}}\hbar \chi_\pm(t))a_{\pm}^{\dagger} \notag\\
        \rho =& \langle\rho_{\chi}\rangle. \notag
    \end{align}
    In this form the equation is almost identical to Eq.~\eqref{eq:2QubitTimeEvolution}
    with the small difference that the thermal noise has been added to the
    deterministic driving force.
    The solution $\rho_{\xi}$ is similar to the zero temperature case, with the
    addition of  the random thermal process. Solution can be
    expressed in terms of coherent states, whose labels satisfy
    \begin{equation}
        \label{eq:noisyTrajectory}
        \dot{z}_{\pm}^{j}+\gamma z_{\pm}^{j} = \dfrac{1}{i\hbar}\left(\widetilde{f}_{\pm}(j) + \sqrt{2\gamma\bar{n}}\hbar\chi_{\pm}\right).
    \end{equation}
    However an important difference is that due to the noise the path can no longer be closed reliably by choosing an appropriate force.
    As can be seen in Fig.~\ref{fig:thermal_traj}, the thermal noise causes
    fluctuations around the paths driven by deterministic force ($z_+^1,\, z_-^0$) and
    the non-driven paths ($z_+^0,\, z_-^1$) are no longer stationary but fluctuate
    around  the ground state.
\subsection{Fidelity in the finite temperature case}
    \label{ssec:fidelityFiniteTemperature}
\begin{figure}
      \includegraphics[width=0.5\textwidth]{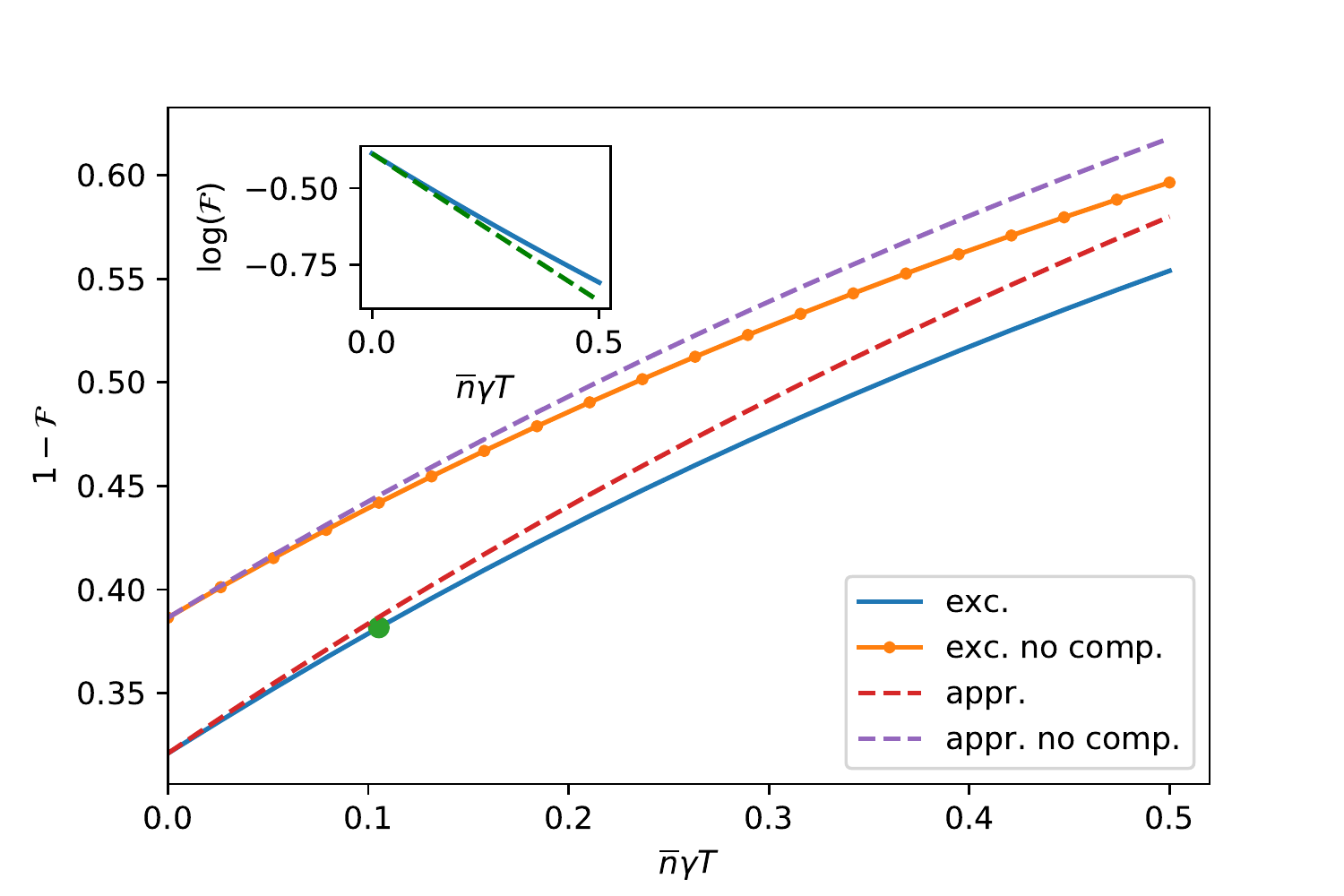}
      \caption{\label{fig:thermal_fid}{(color online) Infidelity with respect to
      thermal fluctuations and the logarithm of fidelity with compensation (inset). Compensated
      force clearly improves the fidelity even in the presence of thermal fluctuations.
      When $\gamma\bar{n}T\approx 0.1$ then
      $1-\mathcal{F}\approx 0.38$ (dot), which corresponds to
      approximately $19\%$ increase compared to the zero temperature.
      Fidelity
      decreases sub-exponentially as a function $\gamma\bar{n}T$, as can be
      seen from the inset. Dashed lines corresponds to a second order
      cumulant expansion of the fidelity. Other parameters are as
      in Fig.~\ref{fig:2QubitAlphaPath} c), except $\gamma/\omega = 0.2$.}}
    \end{figure}

    {One important question is how the fidelity of the phase gate is affected by
      the thermal noise and whether the compensation strategy suggested
      in~\ref{ssec:force} still improves this fidelity.
      In this section we seek to answer this question.\\
      As in section \ref{ssec:fidelity} the fidelity is defined as
    \begin{equation*}
        \mathcal{F} = \bra{\Psi_{\mathrm{d}}}\rho(T)\ket{\Psi_{\mathrm{d}}}.
    \end{equation*}
    Since in the finite temperature case the density operator is given as an average
    $ \rho = \langle\rho_{\chi}\rangle$ the fidelity can be obtained by averaging as well,
    \begin{equation*}
        \mathcal{F} = \langle\bra{\Psi_{\mathrm{d}}}\rho_{\chi}(T)\ket{\Psi_{\mathrm{d}}}\rangle.
    \end{equation*}
    We will proceed to first evaluate the fidelity for a general $\rho_{\chi}(T)$ and
    then take the average over the stochastic processes $\chi_\pm(t) $.\\
    We again consider the overlap with the target state
    $\ket{\Psi_{\mathrm{d}}} = ae^{i\phi_{{\mathrm{isol}}}}\ket{P} + b\ket{A}$ and
    choose $a=b=1/\sqrt{2}$. Other choices for the prefactors will lead to
    higher fidelity as discussed in Sec.~\ref{ssec:fidelity}. The fidelity of
    any $\rho_{\chi}$ after the phase gate operation is given as
    \begin{flalign}
    \label{eq:fidelityFull}
    	\mathcal{F} = &\dfrac{1}{4}( \exp(-(\abs{\zmo}^2+\abs{\zpo}^2))  + \exp(-(\azmz + \azpz))\notag\\
        & +2\operatorname{Re}(e^{-i\Delta\phi - \Gamma -\dfrac{1}{2}\left(\azmo + \azpo + \azmz + \azpz \right)})), \\
        \Gamma = &\gamma\int_{0}^{T} \abs{z_{+}^{1}(\tau) - z_{+}^{0}(\tau)}^{2} + \abs{z_{-}^{1}(\tau) - z_{-}^{0}(\tau)}^{2} \diff{\tau}, \notag\\
        \Delta\phi = &\phi(T) - \phi_{\mathrm{isol}}.\notag
    \end{flalign}
    All quantities in the expression for $\salg$ are evaluated at time $T$, that is,
    at the end of the phase gate operation.\\
    In the expression one can identify the three possible causes for fidelity-loss:
    \begin{enumerate}
    	\item The terms $ \propto \exp(\abs{z_\pm^j(T)}^2) $ arise when the path is not closed, which means that $ z(T) \neq 0 $.
    	\item The term $ \exp(-i\Delta\phi) $ occurs when $ \phi(T) \neq \phi_{\mathrm{isol}} $,
    	\item and the term $ \exp(-\Gamma) $ describes decoherence, which is induced by the damping and can not be prevented.
    \end{enumerate}
    In the zero temperature case we could close the path and restore the phase
    of the isolated case. Therefore only the decoherence was contributing to a
    fidelity loss and the formula reduced to Eq.~\eqref{eq:fidelity}. For finite
    temperature this is no longer the case and we therefore have to work
    with the full expression for $ \mathcal{F} $. We have plotted
    the average infidelity as a function of $\gamma\bar{n}T$ over
    5000 realizations in Fig.~\ref{fig:thermal_fid} (solid line). We checked that our
    result coincide with directly solving the master equation (\ref{eq:model}) (not shown
    in the Figure). The other parameters are as in Fig.~\ref{fig:2QubitAlphaPath} c),
    except $\gamma/\omega=0.2$.  We see that the compensation strategy improves
    the fidelity even at finite temperatures. {The reason is that it still increases the average overlap of the final state and the ground state. }{We note, however, that the importance of this effect decreases
    as $\bar{n}/(\gamma T)$  increases}.
    When $\gamma\bar{n}T\approx 0.1$ (dot) $\mathcal{F}\approx 0.61$ and given the values
    $T=0.3\mu$s, and $\gamma/\omega = 0.2$ and $\omega/2\pi=2 $\, MHz, the
    average thermal photon number $\bar{n}\approx 2.7$.
    This corresponds
    to a temperature of about $0.3$ mK (we evaluate $\bar{n}$ at $\omega$).
    {These temperatures are within the range of current experimental
      capabilities, see for example~\cite{Feldker2020}}.\\
    In the appendix \ref{sec:AnhangFidelity} we evaluate the $\chi$ dependence of the fidelity.
    We use a cumulant expansion to find an approximate analytical expression for the average fidelity.
    This approximation is plotted in Fig.~\ref{fig:thermal_fid} (dashed line)\footnote{For the
    plot we used the results \eqref{eq:avg1stTerm} and \eqref{eq:avg2ndTerm} instead of the final
    linearized expression \eqref{eq:avgFidelity} since those provided better agreement in the rather strongly damped regime}. The inset
    shows that our approximation gives the first order correction in $\gamma\bar{n} T$
    to the zero temperature result. This approximation is motivated by
    the fact that for current ion traps the operation time $T$ is much faster then the coherence
    time $1/(\bar{n}\gamma)$\cite{Schafer2018, doi:10.1063/1.5088164}.
    Our scenario is in the range $\bar{n}\gamma\sim 10^{-1}T$ and
    as we can see, {the approximation nicely captures the finite temperature effects
    on the fidelity in this regime. The rather complicated approximate expression can
    be found in the appendix.}}\\
    \section{Summary and  Outlook}
    \label{sec:summary}
    We examined how phase gates based on the geometrical phases of driven
    trapped ions  behave under dissipation.  We used forces which depend on an
    internal state of the trapped ion in order  to construct relative phases
    which show up in the density operator. We then showed that in the
    special case of a  GKSL-type  evolution with closed phase space paths
    admitting  pure state
    solutions  the total phase will  always have an additional  third
    contribution beyond dynamical- and geometric phases due to dissipation
    \begin{flalign*}
    	\phi &= \phi_{\mathrm{d}} + \phi_{\mathrm{g}} + \xi,\\
    	\xi &= \phi_{\mathrm{L}} + i\eta.
    \end{flalign*}
    This third contribution will in general be {complex} and can be directly
    related to the Lindblad operators. Applied to the harmonic oscillator
    this means that the damping results in a new additional phase that can
    however vanish for certain special cases. More severely, dephasing occurs
    which cannot be avoided and depends on the amplitude of the
    oscillation and the damping strength. We applied our results obtained for a
    single trapped ion to a two-qubit phase gate proposed in
    \cite{PhysRevA.95.022328} which is based on two trapped ions.
    We found that in the presence of damping the phase
    produced by the gate depends on the area swept in the interaction picture alone,
    if the ion is in the ground state at the
    beginning of the operation. However due to the dephasing the fidelity
    of the gate is reduced. This loss of fidelity is especially noticeable
    for large damping strengths or short operation times. Furthermore, our
    calculations show how due to the damping the phase gate no
    longer operates independently of the initial motional state. On
    the other hand it is possible to maintain the robustness against small
    constant offsets of the force $ f \mapsto f + \delta f $ in the damped
    case by constructing the force using  a Gram-Schmidt procedure that also
    ensures that the force produces a closed phase space path  evolution. {We considered
    also finite temperature effects in order to assess the feasibility of this
    scheme. We conclude that this scheme could be soon within reach
    of current experimental techniques by reducing the operation time of
    the gate or by using colder ion traps, thus preventing the onset of thermal fluctuations.}
  
  {To give fair assessment of the relevance of our results, we point out that 
    at the moment two qubit gates are typically experimentally implemented 
    under conditions where the motional state of the ion is rather heated than 
    damped~\cite{PhysRevLett.117.060504,PhysRevLett.117.060505}.
    However, our results predict improvement on the performance of the geometric 
    phase gate if it  would be implemented using buffer gas cooling
    scenario~\cite{Feldker2020} where damping can be the main source for gate errors.}
  
{This work could be extended to more than two ions if the
    block structure of the Hamiltonian is kept and only independent
    decay processes are considered. In future work we will model
    the system from first principles in order to determine the validity of
    the phenomenological model presented and analyzed in this article.}

\acknowledgments
The authors would like to thank Valentin Link for fruitful discussions {and
the anonymous Referees for valuable comments.}
\appendix
\section{Analyzing the phase of a dissipative time evolution}
    \label{sec:Anhangphidissipative}
    Since we assumed that the evolution can be described by a pure state
    we can construct two solutions to Eq.~(\eqref{eq:time evolution}):
    \begin{align*}
    \rho_{00}(t) &= \kb{0}{0}\otimes\kb{\Psi_0(t)}{\Psi_0(t)} \\
    \rho_{11}(t) &= \kb{1}{1}\otimes\kb{\Psi_1(t)}{\Psi_1(t)}.
    \end{align*}
    In order to determine the relative phase between those two states we
    examine the evolution of the superposition which means that we have a
    density operator of the form
    \begin{flalign}
    \rho &= \rho_{00} + \rho _{01} + \left(\rho_{01}\right)^{\dagger} + \rho_{11} \notag\\
    \label{eq:rho01A}
    \rho_{01} &= e^{i\phi(t)} \kb{0}{1}\otimes\kb{\Psi_0}{\Psi_1}.
    \end{flalign}
    We already know that $ \rho_{00} $ and $ \rho_{11} $ solve the
    equation so after inserting $ \rho $ into Eq.~(\ref{eq:time evolution}) we
    are left with
    \begin{align}
    	\label{eq:Anhangrho01dot}
    	\dot{\rho}_{01} &= \frac{-i}{\hbar}[{H},{\rho_{01}}] + \mathcal[\rho_{01}]\\
    	\mathcal{L}[\rho_{01}] &= \sum_{l=1}^{N} L_l\rho_{01}L_l^{\dagger} - \dfrac{1}{2}\left(L_l^{\dagger}L_l\rho_{01} + \rho_{01} L_l^{\dagger}L_l\right)\notag.\\
    \end{align}
    In analogy to \cite{PhysRevLett.58.1593} we can use \eqref{eq:rho01A} and calculate
    \begin{align}
    	\dot{{\rho}}_{01} &= -i\dot{\phi}{\rho}_{01} + e^{-i\phi(t)} \frac{\diff}{\diff t} \left(\kb{0}{1}\otimes\kb{\Psi_0}{\Psi_1}\right)\\
    	-\dot{\phi} &= -i\bra{\Psi_0} \left(\frac{\diff}{\diff t} \kb{\Psi_0}{\Psi_1}\right) \ket{\Psi_1} + ie^{-i\phi}\bra{\Psi_0}\dot{\rho}_{01}\ket{\Psi_1}.
    \end{align}
    Since we want to determine $ \dot{\phi} $ this leaves us with two terms to evaluate:
    \begin{equation*}
    	\bra{{\Psi}_0}\left(\frac{\diff}{\diff t} \kb{\Psi_0}{\Psi_1}\right)\ket{{\Psi}_1} = \langle{{\Psi}_0}\vert{\dot{{\Psi}}_0}\rangle + \langle{\dot{{\Psi}}_1}\vert{{\Psi}_1}\rangle,
    \end{equation*}
    and
    \begin{align*}
    	e^{-i\phi(t)}\bra{{\Psi}_0}\dot{\rho}_{01}\ket{{\Psi}_1} &= \frac{-ie^{-i\phi}}{\hbar}\left( \bra{\Psi_0}[{H},{\rho_{01}}] + L\left[\rho_{01}\right]\ket{\Psi_1} \right)\\
    		&= \dfrac{-i}{\hbar}\left(\bra{\Psi_0}H\ket{\Psi_0} - \bra{\Psi_1}H\ket{\Psi_1}\right) \\
    		&+ \sum_{l=1}^{N}\bra{\Psi_0}L_l\ket{\Psi_0}\bra{\Psi_1}L_l^{\dagger}\ket{\Psi_1} \\
    		&- \dfrac{1}{2}\left(\bra{\Psi_0}L_l^{\dagger}L_l\ket{\Psi_0} + \bra{\Psi_1}L_l^{\dagger}L_l\ket{\Psi_1}\right).
    \end{align*}
    This leads to the final result
    \begin{align*}
    	-\frac{\diff{\phi}}{\diff t} =& -i\left(\bra{\Psi_0}\partial_t\ket{\Psi_0} - \bra{\Psi_1}\partial_t\ket{\Psi_1}\right) +\frac{1}{\hbar}(\langle H_0\rangle - \langle H_1\rangle) \\
    	& i\sum_{l=1}^{N}\bra{\Psi_0}L_l\ket{\Psi_0}\bra{\Psi_1}L_l^{\dagger}\ket{\Psi_1} \\
    	&- \dfrac{1}{2}\left(\bra{\Psi_0}L_l^{\dagger}L_l\ket{\Psi_0} + \bra{\Psi_1}L_l^{\dagger}L_l\ket{\Psi_1}\right).
    \end{align*}
    Here we also used that $ \langle{\Psi}\vert\dot{\Psi}\rangle $ is purely
    imaginary and therefore
    $ \langle{\Psi}\vert{\dot{\Psi}}\rangle = -\langle{\dot{\Psi}}\vert{\Psi}\rangle $.
{\section{Fidelity at finite temperature}}
\label{sec:AnhangFidelity}
{In Sec.~\ref{sec:finiteTemperatureEffects} we derived an expression
for the fidelity of the gate for finite temperatures. Since this expression depends on
the random noise $ \chi $ we have to take the average of $\mathcal{F}$ over $\chi$ in order to make
meaningful predictions about the fidelity of the phase gate. \\
We will first take the averages of the terms $\propto\exp(-\abs{z_{\pm}^{j}(T)}^2)$ and later
evaluate the average of the term $ \exp(-i\Delta\phi - \Gamma -\dfrac{1}{2}\left(\azmo + \azpo + \azmz + \azpz \right)) $.\\
  In the finite temperature case the presence of noise leads to the following equation for $ z(t) $:
  \begin{equation}
    \label{eq:z(t)FiniteT}
	z_{\pm}^{j}(t) = \dfrac{1}{i\hbar}\int_{0}^{t}(f_{\pm}^{j}+ \sqrt{2\gamma\bar{n}}\hbar\chi_{\pm})e^{i\Omega_{\pm}\tau - \gamma (t-\tau)}\diff{\tau}.
  \end{equation}
  Note that there are two different uncorrelated noises $ \chi_{\pm} $ for the two
different modes of oscillation (stretch and center of mass mode) that do however
not depend on the internal states. Since the noise is just added to the force
we can split $ z $ in two parts, where one part $ z_{\pm,f}^{j}(t) $ is under full
control of the force (zero temperature path) and the second part $ z_{\pm,\chi}^{j}(t) $ is a fluctuation due to the noise.
  \begin{flalign}
	\label{eq:splitZ}
	z_{\pm}^{j}(t) &= z_{\pm,f}^{j}(t)+z_{\pm,\chi}^{j}(t),\\
	z_{\pm,f}^{j}(t) &= \dfrac{1}{i\hbar}\int_{0}^{t}f_{\pm}^{j}e^{i\Omega_{\pm}\tau - \gamma (t-\tau)}\diff{\tau}, \notag\\
	z_{\pm,\chi}^{j}(t) &= \dfrac{1}{i\hbar}\int_{0}^{t} \sqrt{2\gamma\bar{n}}\hbar\chi_{\pm}e^{i\Omega_{\pm}\tau - \gamma (t-\tau)}\diff{\tau}.\notag
  \end{flalign}
  Since $z_{\pm,f}^{j}$ does not depend on $\chi$ we arrive at
  \begin{flalign}
    \label{eq:fidelityAbsTerms}
	&\langle\exp(-\abs{z_{\pm}^{j}(T)}^2)\rangle = e^{-\abs{z_{\pm,f}^{j}(T)}^2} \notag\\
	&\times \langle\exp\Big[-z_{\pm,f}^{j*}(T)z_{\pm,\chi}^{j}(T)
	- z_{\pm,f}^{j}(T)z_{\pm,\chi}^{j*}(T) \notag\\
	&\hspace{20pt}- \abs{z_{\pm,\chi}^{j}(T)}^2\Big]\rangle
  \end{flalign}
  If we define
  \begin{flalign}
    \mathcal{V}_{\pm}^{j}(\tau) &= -z_{\pm,f}^{j*}(T)\dfrac{1}{i\hbar}\sqrt{2\gamma\bar{n}}\hbar\chi_{\pm}e^{i\Omega_{\pm}\tau - \gamma(T-\tau)} \notag\\
    &- z_{\pm,f}^{j}(T)\dfrac{1}{-i\hbar}\sqrt{2\gamma\bar{n}}\hbar\chi_{\pm}^{*}e^{-i\Omega_{\pm}\tau - \gamma(T-\tau)}\notag\\
    &-  \int_{0}^{T} 2\gamma\bar{n}\chi_\pm(\tau)\chi^{*}_\pm(s) e^{i\Omega_\pm(\tau-s)-\gamma(T-\tau + T -s)}\diff{s},
  \end{flalign}
  we can rewrite Eq.~\eqref{eq:fidelityAbsTerms} as
  \begin{flalign}
\label{eq:fidelityV}
\langle e^{-\abs{z_{\pm}^{j}(T)}^2}\rangle &= e^{-\abs{z_{\pm,f}^{j}(T)}^2}
\langle e^{\int_{0}^{T} \mathcal{V}_{\pm}^{j}(\tau) \diff{\tau}}\rangle .
  \end{flalign}
  We now evaluate the average term by taking the cumulant expansion to 2nd order:
  \begin{flalign*}
    \langle e^{\int_{0}^{T} \mathcal{V}_{\pm}^{j}(t) \diff{t}}\rangle = e^{\int_{0}^{T} \langle\mathcal{V}_{\pm}^{j}(t)\rangle \diff{t}
      + \dfrac{1}{2} \int_{0}^{T}\int_{0}^{T} \langle\mathcal{V}_{\pm}^{j}(t_1)\mathcal{V}_{\pm}^{j}(t_2)\rangle\diff{t_1} \diff{t_2} + \dots}
  \end{flalign*}
  Using the Gaussian nature of the processes $\chi_\pm$ we get
  \begin{flalign}
    \langle e^{-\abs{z_{\pm}^{j}(T)}^2}\rangle\approx& \exp\Big[-\bar{n}\left(1-e^{-2\gamma T}\right) \notag\\
    &-\abs{z_{\pm,f}^{j}}^{2}(1-\bar{n}\left(1-e^{-2\gamma T}\right)) \notag\\
    	\label{eq:avg1stTerm}
    	&+  \dfrac{3}{2} \left(\bar{n}\left(1-e^{-2\gamma T}\right)\right)^{2}\Big] .
      \end{flalign}
      From the result we can see that the finite temperature leads to additional decoherence
      which increases even further for $ z_{\pm,f}^{j}(T) \neq 0 $. Note that $ z_{\pm,f}^{j}(T) = 0 $
      when the path is closed in the zero temperature case. The
      third term in the expansion above is of the order $ (\gamma\bar{n}T)^{2} = (T/\tau_{\mathrm{d}})^{2} $.
      Since the coherence time $ \tau_{\mathrm{d}} $ is much larger than the
      operation time for current ion-traps \cite{Schafer2018, lucas2007longlived}
      we will only keep terms in the first order of $ \gamma\bar{n}T $ from here on.\\
      We are now going to take a look at the average of the term
      \begin{equation}
        \label{eq:2ndTerm}
        \langle e^{-i\Delta\phi - \Gamma -\dfrac{1}{2}\left(\azmo + \azpo + \azmz + \azpz \right)}\rangle.
      \end{equation}
      At first one finds that $ \Gamma $ is actually independent of $ \chi $ since it is
      defined as the difference between two paths of different internal states
      (see Eq.~\eqref{eq:fidelityFull}) and since the noise is independent
      of the internal state it cancels out.
      We than follow the same strategy as before and introduce a $\mathcal{V}'(t)$
      and approximate the exponential with a cumulant expansion
      \begin{equation*}
    	\mathcal{V}'(\tau) = -i\dot{\phi_{\chi}}(\tau) +\dfrac{1}{2}\left(\mathcal{V}_{-}^{1}(\tau) + \mathcal{V}_{+}^{1}(\tau)+ \mathcal{V}_{-}^{0}(\tau)+\mathcal{V}_{+}^{0}(\tau)\right) .
    \end{equation*}
    Like the trajectory before we also split the phase into a part that is determined by the force
    alone $ \phi_{f} $ and a part that depends on the noise $ \phi_{\chi} $: $ \phi = \phi_f + \phi_{\chi} $
	With this definition we can write Eq.~\eqref{eq:2ndTerm} as
	\begin{flalign*}
      e^{-i(\phi_f - \phi_{\mathrm{isol}}) - \Gamma -\dfrac{1}{2}\left(\abs{z_{f,-}^{1}}^{2}
          +\abs{z_{f,+}^{1}}^{2}+\abs{z_{f,-}^{0}}^{2}+\abs{z_{f,+}^{0}}^{2} \right)}
      \langle e^{\int_{0}^{T} \mathcal{V}'(\tau) \diff{\tau}}\rangle .
	\end{flalign*}
	We again take the cumulant expansion up to 2nd order and only keep terms up to first order of $ \gamma\bar{n}T $ to arrive at:
	\begin{widetext}
      \begin{flalign}
        \label{eq:avg2ndTerm}
        &\langle \exp({-i\Delta\phi - \Gamma -\frac{\azmo + \azpo + \azmz + \azpz}{2}})\rangle
         \approxeq \exp\Bigg[ -i(\phi_f(T) - \phi_{\mathrm{isol}}) - \Gamma -2\bar{n}\left({1-e^{-2\gamma T}}\right) \notag\\
		 &\hspace{10pt}+ \dfrac{\gamma\bar{n}}{\hbar}\operatorname{Im}\left(\int_{0}^{T}\diff{t_1}\int_{0}^{t_1}\diff{t_2} \left(z_{-,f}^{1*}(t_2)\widetilde{f}_{-}^{1}(t_1) + z_{+,f}^{0*}(t_2)\widetilde{f}_{+}^{0}(t_1) \right)e^{-\gamma(T+t_1 - 2t_2)} \right) \notag\\
        &\hspace{10pt}-\dfrac{\bar{n}}{\hbar^2}\operatorname{Re}\left(\int_{0}^{T}\diff{t_1}\int_{0}^{t_1}\diff{t_2} \left(\widetilde{f}_{-}^{1}(t_1)\widetilde{f}_{-}^{1*}(t_2) +
          \widetilde{f}_{+}^{0}(t_1)\widetilde{f}_{+}^{0*}(t_2)\right)e^{-\gamma t_1}\sinh(\gamma t_2)\right)\notag\\
		&\hspace{10pt}+ \dfrac{\gamma\bar{n}}{2}\int_{0}^{T}\diff{t_1}\abs{z_{-,f}^{1}(t_1)}^{2} + \abs{z_{+,f}^{0}(t_1)}^{2}
		+ \dfrac{i\bar{n}}{\hbar}\operatorname{Re}\left(\int_{0}^{T}\diff{t_1}\left(z_{-,f}^{1}(T)\widetilde{f}_{-}^{1}(t_1) - z_{+,f}^{0}(T)\widetilde{f}_{+}^{0}(t_1)\right)e^{-\gamma T}\sinh(\gamma t_1)\right)\notag\\
		&\hspace{10pt}-\dfrac{\abs{z_{-,f}^{1}(T)}^{2}+ \abs{z_{+,f}^{0}(T)}^{2}}{2}{\left(1-\bar{n}\left(1-e^{-2\gamma T}\right)\right)}\Bigg] .
       \end{flalign}
       With Eq.~\eqref{eq:avg1stTerm}, \eqref{eq:avg2ndTerm} we can now express the average fidelity in first order of $ \bar{n}\gamma T $, where we also expanded the exponentials of $\gamma T$:
       \begin{flalign}\label{eq:avgFidelity}
         \langle\mathcal{F}\rangle \approxeq& \dfrac{\exp(-4\bar{n}\gamma T - \abs{z_{-,f}^{1}(T)}^{2}(1-2\bar{n}\gamma T))+\exp(-4\bar{n}\gamma T - \abs{z_{+,f}^{0}(T)}^{2}(1-2\bar{n}\gamma T))}{4} \notag\\
	     &+ \frac{1}{2}\operatorname{Re}\Bigg[ \exp\Bigg\{ -i(\phi_f(T) - \phi_{\mathrm{isol}}) - \Gamma -4\bar{n}\gamma T - \dfrac{\gamma\bar{n}}{2}\int_{0}^{T}\diff{t_1}\abs{z_{-,f}^{1}(t_1)}^{2} + \abs{z_{+,f}^{0}(t_1)}^{2} \notag\\
	     &+ \dfrac{\gamma\bar{n}}{\hbar}\operatorname{Im}\left(\int_{0}^{T}\diff{t_1}\int_{0}^{t_1}\diff{t_2} z_{+,f}^{0*}(t_2)\widetilde{f}_{+}^{0}(t_1)+z_{-,f}^{1*}(t_2)\widetilde{f}_{-}^{1}(t_1)\right) \notag\\
	     &+ \dfrac{i{\gamma}\bar{n}}{\hbar}\operatorname{Re}\left(\int_{0}^{T}\diff{t_1}\left(z_{-,f}^{1}(T)\widetilde{f}_{-}^{1}(t_1) - z_{+,f}^{0}(T)\widetilde{f}_{+}^{0}(t_1)\right){t_1}\right)\notag\\
         &{-}\dfrac{{\gamma}\bar{n}}{\hbar^2}{\operatorname{Re}}\left(\int_{0}^{T}\diff{t_1}\int_{0}^{t_1}\diff{t_2} \left(\widetilde{f}_{-}^{1}(t_1)\widetilde{f}_{-}^{1*}(t_2) +
           \widetilde{f}_{+}^{0}(t_1)\widetilde{f}_{+}^{0*}(t_2)\right){t_2}\right)\notag\\
	     &-\dfrac{\abs{z_{-,f}^{1}(T)}^{2}+ \abs{z_{+,f}^{0}(T)}^{2}}{2}\left(1-{2}\bar{n}\gamma T\right)\Bigg\}\Bigg] .
      \end{flalign}
      {Note that this expression still contains the endpositions of the zero temperature paths $\abs{z_{\pm,f}^{j}(T)}^{2}$. Since these vanish if the compensation strategy is employed, this suggests that the strategy improves the fidelity even at finite temperatures. Furthermore, although all of the terms in the exponential are of first order of $\gamma\bar{n}T$ we found that for our set of forces the dominating temperature dependent contribution would be the $-4\bar{n}\gamma T$ terms. They arise because the finite temperature paths can no longer be closed reliably.}
	\end{widetext}}

\bibliography{literature}
\end{document}